\documentclass[preprintnumbers,amsmath,amssymb]{revtex4}

\usepackage{epsfig}
\usepackage{graphicx}
\usepackage{dcolumn}
\usepackage{bm}
\usepackage{tikz}
\usetikzlibrary{arrows}
\usepackage[all]{xy}
\usepackage{url}

\usepackage{color} 
\usepackage{soul} 

\usepackage{fancyheadings} 
\fancypagestyle{first}{%
\fancyhf{} 
\cfoot{
\footnotesize{
This material is based upon work supported by Defense Advanced Research Projects Agency (DARPA) and SPAWAR Systems Center Pacific (SSC Pacific) under Contract No. N66001-13-C-4047. The views, opinions, and/or findings contained in this article/presentation are those of the author(s)/presenter(s) and should not be interpreted as representing the official views or policies of the Department of Defense or the U.S. Government. Distribution Statement ``A'' (Approved for Public Release; Distribution Unlimited).
}
} 
 
} 

\newcommand{\tstamp}{\today}   
\begin{document}

\title{De Bruijn entropy and string similarity}

\author{Steve Huntsman}%
\email{steve.huntsman@baesystems.com}
\affiliation{BAE Systems, 4301 North Fairfax Drive, Arlington, Virginia 22203
}%

\author{Arman Rezaee}%
\email{armanr@mit.edu}
\affiliation{MIT EECS, 77 Massachusetts Avenue, Cambridge, Massachusetts 02139
}%

%

\date{\today}

\begin{abstract}
We introduce the notion of \emph{de Bruijn entropy} of an Eulerian quiver and show how the corresponding relative entropy can be applied to practical string similarity problems. This approach explicitly links the combinatorial and information-theoretical properties of words and its performance is superior to edit distances in many respects and competitive in most others. The computational complexity of our current implementation is parametrically tunable between linear and cubic, and we outline how an optimized linear algebra subroutine can reduce the cubic complexity to approximately linear. Numerous examples are provided, including a realistic application to molecular phylogenetics.
\end{abstract}


\maketitle


\thispagestyle{first}

\section{\label{sec:Introduction}Introduction}

String similarity is a fundamental problem touching on computer science, bioinformatics, machine learning, and many other areas \cite{Rieck}. Most fast approaches to string similarity (e.g., bag-of-words or kernel methods) are heuristic, whereas most theoretically grounded approaches to string similarity (e.g., Kolmogorov complexity methods) are slow. In this paper, we discuss a technique that bridges the gap, offering performance that can be tuned between linear and (in a sufficiently optimized implementation) subquadratic time while offering a clear interpretation in terms of combinatorial and information-theoretical primitives. Our technique is particularly well suited for comparing words based on their local structure and is agnostic to global structure, which is particularly interesting for comparing strings/words encoding paths through digraphs with cycles (e.g., control flow graphs of computer programs) or for streaming data. 

The paper is structured as follows. We begin by establishing notation and graph constructions in \S \ref{sec:Preliminaries} before discussing basic combinatorial properties in \S \ref{sec:Combinatorics} and our ultimate information-theoretical considerations in \S \ref{sec:InformationTheory}. Finally, we outline applications, focusing in particular on molecular phylogeny as an example where an approximate ``ground truth'' furnishes a basis for evaluating the performance of our approach and its comparison with conventional techniques. Appendices on spin models as well as code for reproducing and extending our results are included.

\section{\label{sec:Preliminaries}Preliminaries}

We begin with some preliminaries to establish basic definitions and notation. Let $n < \infty$ and consider a finite set $\mathcal{A} := \{a_1,\dots,a_n\}$ which we call an \emph{alphabet}. A \emph{word} or \emph{string} over $\mathcal{A}$ of length $\ell$ is an element of $\mathcal{A}^\ell$; a \emph{symbol} is a word of length 1. The word $w = (w_1,\dots,w_\ell)$ will typically be written as $w = w_1 \dots w_\ell$. With a slight abuse of notation, we write $\ell(w) = \ell$. The \emph{concatenation} of two words $w = w_1 \dots w_\ell$ and $w' = w'_1 \dots w'_{\ell'}$ is $ww' := w_1 \dots w_\ell w'_1 \dots w'_{\ell'}$. 

A \emph{cyclic word} or \emph{necklace} \cite{Stanley} of length $\ell$ is the set of cyclic shifts of a word. We shall engage in a minor abuse of notation by letting $w$ denote either a word or a cyclic word depending on context. If $w$ is cyclic, $w_j := w_{((j-1) \text{ mod } \ell ) + 1}$.

Recall that a \emph{quiver} (also known as a \emph{multidigraph}, \emph{directed multigraph}, etc.) $Q$ is an ordered pair $(V(Q),E(Q)) \equiv (V,E)$ s.t. $E$ is a multiset over $V \times V$ \cite{Bollobas}. The \emph{adjacency matrix} $A(Q)$ of $Q$ is defined so that if there are $a$ edges from $v_j$ to $v_k$, then $A(Q)_{jk} := a$. It is clear that a quiver may be reconstructed from its adjacency matrix and \emph{vice versa}, so that we may write $f(Q) \equiv f(A)$ for a generic function $f$ without any ambiguity so long as either side is defined. Furthermore, we may make the implicit identifications $v_j \equiv j$ and $Q \equiv A(Q)$ for convenience.

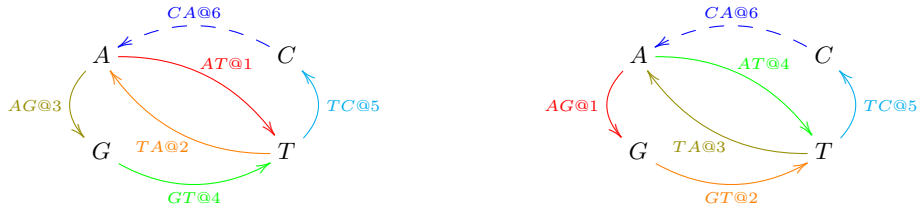
\begin{figure}[htbp]
\begin{equation}
\xymatrix@C=2cm{
A \ar@/^-1pc/@[olive][d]_{{\color{olive}AG@3}} \ar@/^1pc/@[red][dr]^{{\color{red}AT@1}}
& C \ar@/^-1pc/@{-->}@[blue][l]_{{\color{blue}CA@6}} 
&
& A \ar@/^-1pc/@[red][d]_{{\color{red}AG@1}} \ar@/^1pc/@[green][dr]^{{\color{green}AT@4}}
& C \ar@/^-1pc/@{-->}@[blue][l]_{{\color{blue}CA@6}}
\\
G \ar@/^-1pc/@[green][r]_{{\color{green}GT@4}}
& T \ar@/^1pc/@[orange][ul]^{{\color{orange}TA@2}} \ar@/^-1pc/@[cyan][u]_{{\color{cyan}TC@5}}
&
& G \ar@/^-1pc/@[orange][r]_{{\color{orange}GT@2}}
& T \ar@/^1pc/@[olive][ul]^{{\color{olive}TA@3}} \ar@/^-1pc/@[cyan][u]_{{\color{cyan}TC@5}}
}
\nonumber
\end{equation}
\caption{ \label{fig:ATAGTCandAGTATC} The cyclic words $ATAGTC$ (L) and $AGTATC$ (R) have identical order 1 de Bruijn quivers and (equivalently) the same 2-grams over the alphabet $\{A,C,G,T\}$. The quiver edges are annotated with $w_j w_{j+1}@j$ and colored according to $j$. Removing the dashed edges yields quivers for non-cyclic words in the obvious way.} 
\end{figure}

For $w$ cyclic and $k < \ell(w)$, the \emph{order $k$ de Bruijn quiver}
\footnote{
NB. Similar constructions (usually digraphs rather than quivers) appear throughout the literature: see e.g., \cite{Moreno} for a recent example. 
}
$Q_k(w)$ is given by
\begin{itemize}
\item $V(Q_k(w)) := \mathcal{A}^k$;
\item $E(Q_k(w)) := \{ (w_{1+j} \dots w_{k+j}, w_{2+j} \dots w_{k+1+j}) : 0 \le j < \ell \}$.
\end{itemize}
That is, the edges of $Q_k(w)$ correspond to the subwords of length $(k+1)$ (a/k/a \emph{$(k+1)$-grams}) of $w$, with multiplicities counted. Figure \ref{fig:ATAGTCandAGTATC} shows an example.

\textsc{Remarks.}
\begin{itemize}
\item The order 0 de Bruijn quiver of a word $w$ has one vertex corresponding to the empty word and edges corresponding to each symbol in $w$, with multiplicity.
\item If $w$ is a $n$-ary de Bruijn sequence of length $n^k$, then $Q_{k-1}(w)$ is the $n$-ary de Bruijn graph with $n^k$ edges. 
\item $Q_k(w)$ is \emph{Eulerian} (i.e., [strongly] connected and with indegrees equal to outdegrees, so that we may unambiguously write $\deg(v)$ for either quantity at vertex $v$) iff $w$ contains every possible $k$-gram (otherwise, there are isolated vertices, but we may elide this technicality without comment at times). An Euler circuit on $Q_k(w)$ corresponds to a Hamiltonian path on $Q_{k+1}(w)$. These properties are why we deal with cyclic words.
\end{itemize}

\section{\label{sec:Combinatorics}Combinatorics}

To begin this section, we remark that it is a concrete application (though perhaps of sufficient generality as to border on a reformulation) of the so-called \emph{transfer matrix} method \cite{Stanley}.

Write $w \sim_k w'$ iff $Q_k(w) = Q_k(w')$. It is clear that $\sim_k$ is an equivalence relation; denote the corresponding equivalence class of $w$ by $[w]_k$. Let $W_k(w) := |[w]_k|$ denote the number of cyclic words with the same order $k$ de Bruijn quiver as $w$. In order to compute $W_k(w)$ it is convenient to consider the adjacency matrix $A_k(w) \equiv A(Q_k(w))$.

If $A$ is a square matrix, write $d(A)$ for the vector with components given by the diagonal entries of $A$. If now $1$ denotes a vector of ones, then $L(A) := d(A1) - A$ is the \emph{Laplacian} of $A$. We recall the following two classical theorems:

\

\textsc{Matrix-tree theorem.} Let $Q$ be a quiver. The diagonal cofactors of $L(A(Q))$ are all equal to each other and to the number $t(Q)$ of directed spanning trees of $Q$ oriented towards any fixed vertex. $\Box$ 

\

\textsc{BEST [de Bruijn, van Aardenne-Ehrenfest, Smith, and Tutte] theorem.} Let $Q$ be an Eulerian quiver. Then the number $c(Q)$ of Euler circuits of $Q$ is
\begin{equation}
\label{eq:BEST}
c(Q) = t(Q) \cdot \prod_{v \in V(Q)} (\deg(v)-1)!. \quad \Box
\end{equation}

\

These readily yield the following 

\

\textsc{Corollary.} \cite{Jonsson,FredricksenHuntsman} Let $A := A_k(w)$ correspond to an Eulerian de Bruijn quiver. Then
\begin{equation}
\label{eq:Ww}
W_k(w) = W(A) := \sum_{d|\gcd(A)} \frac{\phi(d) \cdot c(A/d)}{d \cdot (A/d)!}.
\end{equation}
Here $\phi(\cdot)$ is the totient function, the gcd is defined elementwise and $M! := \prod_{i,j} M_{ij}!$. The $d = 1$ term dominates, giving the simple and effective approximation $W(A) \approx c(A)/A!$. We note that if $Q$ is an Eulerian (not necessarily de Bruijn) quiver with adjacency matrix $A$, then we may still write $W(A)$ or $W(Q)$ for the RHS of \eqref{eq:Ww}. However, it is not necessary to directly interpret $W$ in this more abstract context, since any finite Eulerian quiver can be embedded in some de Bruijn quiver.

\

\textsc{Sketch of proof.} The formula is obvious when $\gcd(A) = 1$, as in this case every cyclic word in $[w]_k$ corresponds to $A!$ Euler circuits. More generally, for $d | m | \gcd(A)$, the term $\frac{\phi(d) \cdot c(A/d)}{d \cdot (A/d)!}$ counts the cyclic words in $[w]_k$ of period $\ell/m$ with multiplicity $\frac{\phi(d)}{d} \cdot \frac{1}{m/d} = \frac{\phi(d)}{m}$. Since $\frac{1}{m} \sum_{d|m} \phi(d) = 1$, the result follows. $\Box$


\subsection{\label{sec:example1}Example 1}

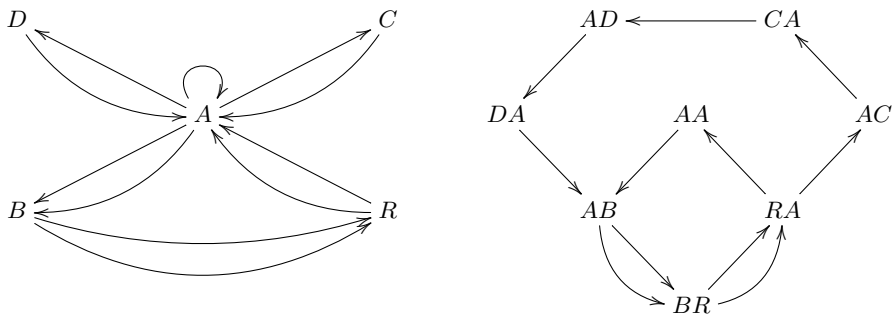
\begin{figure}[htbp]
\begin{equation}
\xymatrix@C=2cm{
D \ar@/^-1pc/[rd]
&
& C \ar@/^1pc/[ld] \\
& A \ar@/^1pc/[ld] \ar[ld] \ar@(lu,ru)[] \ar[lu] \ar[ru]
& \\
B \ar@/^-1pc/[rr] \ar@/^-2pc/[rr]
&
& R \ar[lu] \ar@/^1pc/[lu] 
}
\quad \quad \quad
\xymatrix@C=5mm{
& AD \ar[ld] 
&
& CA \ar[ll]
& \\
DA \ar[rd]
&
& AA \ar[ld]
&
& AC \ar[lu] \\
& AB \ar[rd] \ar@/^-1pc/[rd]
&
& RA \ar[lu] \ar[ru]
& \\
& & BR \ar[ru] \ar@/^-1pc/[ru]
& & \\
}
\nonumber
\end{equation}
\caption{ \label{fig:ABRACADABRA} (L) $Q_1(w)$. (R) $Q_2(w)$. } 
\end{figure}

Consider $w = ABRACADABRA$ over $\mathcal{A} = \{A,B,C,D,R\}$. $Q_1(w)$ and $Q_2(w)$ are depicted in figure \ref{fig:ABRACADABRA}.

We have that 
\begin{equation}
A_1(w) = \begin{pmatrix}
1_{AA} & 2_{AB} & 1_{AC} & 1_{AD} & 0_{AR} \\
0_{BA} & 0_{BB} & 0_{BC} & 0_{BD} & 2_{BR} \\
1_{CA} & 0_{CB} & 0_{CC} & 0_{CD} & 0_{CR} \\
1_{DA} & 0_{DB} & 0_{DC} & 0_{DD} & 0_{DR} \\
2_{RA} & 0_{RB} & 0_{RC} & 0_{RD} & 0_{RR}
\end{pmatrix}
\end{equation}
where for convenience we have annotated matrix entries with subscripts corresponding to the 2-grams. It is easy to check that $t(Q_1(w)) = 4$. Meanwhile, the ordered tuple of vertex degrees (i.e., the row or column sums of $A$) is $(5,2,1,1,2)$, so $\prod_v (\deg(v)-1)! = 24$. Therefore $c(Q_1(w)) = 4 \cdot 24 = 96$. The sum in equation \eqref{eq:Ww} has only a single term, corresponding to $d = 1$, and so $W_1(w) = (\phi(1) \cdot 96) / (1 \cdot (2!)^3 \cdot (1!)^5 \cdot (0!)^{17}) = 96/8 = 12$.
\footnote{
Our MATLAB code readily computes this as \texttt{exp(cequiverentropy(radixwordquiver('ABRACADABRA',1,'ABCDR')))}. Similarly, using \texttt{exp(cequiverentropy(radixwordquiver('ATAGTC',1,'ACGT')))} gives a result of 2.
}
By noting the cycle structure of $Q_1(w)$, we can list the 12 elements of $[w]_1$ by hand. In the lexicographical order inherited from the usual order on $\mathcal{A}$, these are: 
\begin{center}
\begin{tabular}{c | c c c c c c c c c c c }
  1 & $A$ & $B$ & $R$ & $A$ & $B$ & $R$ & $A$ & $C$ & $A$ & $D$ & $A$ \\
  2 & $A$ & $B$ & $R$ & $A$ & $B$ & $R$ & $A$ & $D$ & $A$ & $C$ & $A$ \\
  3 & $A$ & $B$ & $R$ & $A$ & $C$ & $A$ & $B$ & $R$ & $A$ & $D$ & $A$ \\
  4 & $A$ & $B$ & $R$ & $A$ & $C$ & $A$ & $D$ & $A$ & $B$ & $R$ & $A$ \\
  5 & $A$ & $B$ & $R$ & $A$ & $D$ & $A$ & $B$ & $R$ & $A$ & $C$ & $A$ \\
  6 & $A$ & $B$ & $R$ & $A$ & $D$ & $A$ & $C$ & $A$ & $B$ & $R$ & $A$ \\
  7 & $A$ & $C$ & $A$ & $B$ & $R$ & $A$ & $B$ & $R$ & $A$ & $D$ & $A$ \\
  8 & $A$ & $C$ & $A$ & $B$ & $R$ & $A$ & $D$ & $A$ & $B$ & $R$ & $A$ \\
  9 & $A$ & $C$ & $A$ & $D$ & $A$ & $B$ & $R$ & $A$ & $B$ & $R$ & $A$ \\
  10 & $A$ & $D$ & $A$ & $B$ & $R$ & $A$ & $B$ & $R$ & $A$ & $C$ & $A$ \\
  11 & $A$ & $D$ & $A$ & $B$ & $R$ & $A$ & $C$ & $A$ & $B$ & $R$ & $A$ \\
  12 & $A$ & $D$ & $A$ & $C$ & $A$ & $B$ & $R$ & $A$ & $B$ & $R$ & $A$ \\
\end{tabular}
\end{center}

It is not difficult to similarly show that $W_2(w) = 1$. 

Finally, consider $w' = BARBARA$. This is a degenerate case as the symbols $C$ and $D$ are not present, so that the de Bruijn quiver is only \emph{componentwise Eulerian}: i.e., the in- and outdegrees coincide, but some are zero, so that the quiver is not connected. Taking $\mathcal{A}' = \{A,B,R\}$ (or equivalently and perhaps more straightforwardly, modifying the Laplacian by changing any zeros along the diagonal to ones) to remedy this, we get that $t(Q_1(w')) = 4$. The ordered tuple of vertex degrees is $(3,2,2)$, so $\prod_v (\deg(v)-1)! = 2$ and $c(Q_1(w')) = 2 \cdot 4 = 8$. As before, there is a single term in the sum for $W_1(w') = (\phi(1) \cdot 8) / (1 \cdot (2!)^2 \cdot (1!)^3 \cdot (0!)^4) = 2$. The other element of $[w']_1$ is $BARARBA$. $\Box$

\subsection{\label{sec:example2}Example 2}

Consider $\mathcal{A} = \{0,1\}$, $k = 1$, and fix $\ell$. If $g \in \{00,01,10,11\}$, let $x_g(w)$ be the number of times that $g$ occurs in $w$. Because $w$ is cyclic, we must have $x_{01} = x_{10} = (\ell - x_{00} - x_{11})/2 =: x_*$, and $A_1(w) = \left ( \begin{smallmatrix} x_{00} & x_* \\ x_* & x_{11} \end{smallmatrix} \right )$. 
\footnote{
The degenerate case $x_* = 0$ corresponds to the words $0^\ell$ and $1^\ell$, and must be treated separately.
}
We have that $L(A_1(w)) = x_* \left ( \begin{smallmatrix} 1 & -1 \\ -1 & 1 \end{smallmatrix} \right )$, so $t(Q_1(w)) = x_*$. Furthermore, $\deg(0) = x_{00} + x_*$ and $\deg(1) = x_* + x_{11}$, so $c(Q_1(w)) = x_* \cdot (x_{00} + x_* - 1)! \cdot (x_* + x_{11} - 1)!$. It follows after a line or two of algebra that
\begin{equation}
\label{eq:example2eq}
W_1(w) \equiv W_1(x_{00},x_*;\ell) = \frac{x_*}{(x_{00}+x_*)(x_*+x_{11})} \cdot \sum_{d|\gcd(x_{00},x_{11},x_*)} \phi(d) \cdot \binom{(x_{00}+x_*)/d}{x_*/d} \binom{(x_*+x_{11})/d}{x_*/d}.
\end{equation}

	%

Explicitly, for $\ell = 16$, we have the following table of values, with zeros omitted:
\footnote{
This can readily be computed using the following command sequence:
\texttt{
L = 16; for j = 1:L/2, for k = 1:(L-1), A = [k-1,j;j,L-(k-1)-2*j]; ce16(j,k) = cequiverentropy(A); end, end, W = exp(ce16)
}
}
\begin{center}
\begin{tabular}{c c}
& $x_{00}$ \\
$x_*$ &
	\begin{tabular}{c | *{17}{p{8mm}}}
	$W_1$ & 0 & 1 & 2 & 3 & 4 & 5 & 6 & 7 & 8 & 9 & 10 & 11 & 12 & 13 & 14 & 15 & 16 \\
	\hline
	0 & 1 & & & & & & & & & & & & & & & & 1 \\
	1 & 1 & 1 & 1 & 1 & 1 & 1 & 1 & 1 & 1 & 1 & 1 & 1 & 1 & 1 & 1 & & \\
	2 & 7 & 12 & 17 & 20 & 23 & 24 & 25 & 24 & 23 & 20 & 17 & 12 & 7 & & & & \\
	3 & 22 & 55 & 90 & 120 & 140 & 147 & 140 & 120 & 90 & 55 & 22 & & & & & & \\
	4 & 43 & 120 & 212 & 280 & 309 & 280 & 212 & 120 & 43 & & & & & & & & \\
	5 & 42 & 126 & 210 & 245 & 210 & 126 & 42 & & & & & & & & & & \\
	6 & 22 & 56 & 75 & 56 & 22 & & & & & & & & & & & & \\
	7 & 4 & 7 & 4 & & & & & & & & & & & & & & \\
	8 & 1 & & & & & & & & & & & & & & & & \\
	\end{tabular}
\end{tabular}
\end{center}
Summing over the table entries shows that there are 4116 distinct cyclic binary words of length 16, a fact which can be confirmed via the Cauchy-Frobenius lemma.

\begin{figure}[htbp]
\includegraphics[trim = 5mm 0mm 10mm 0mm, clip, width=80mm,keepaspectratio]{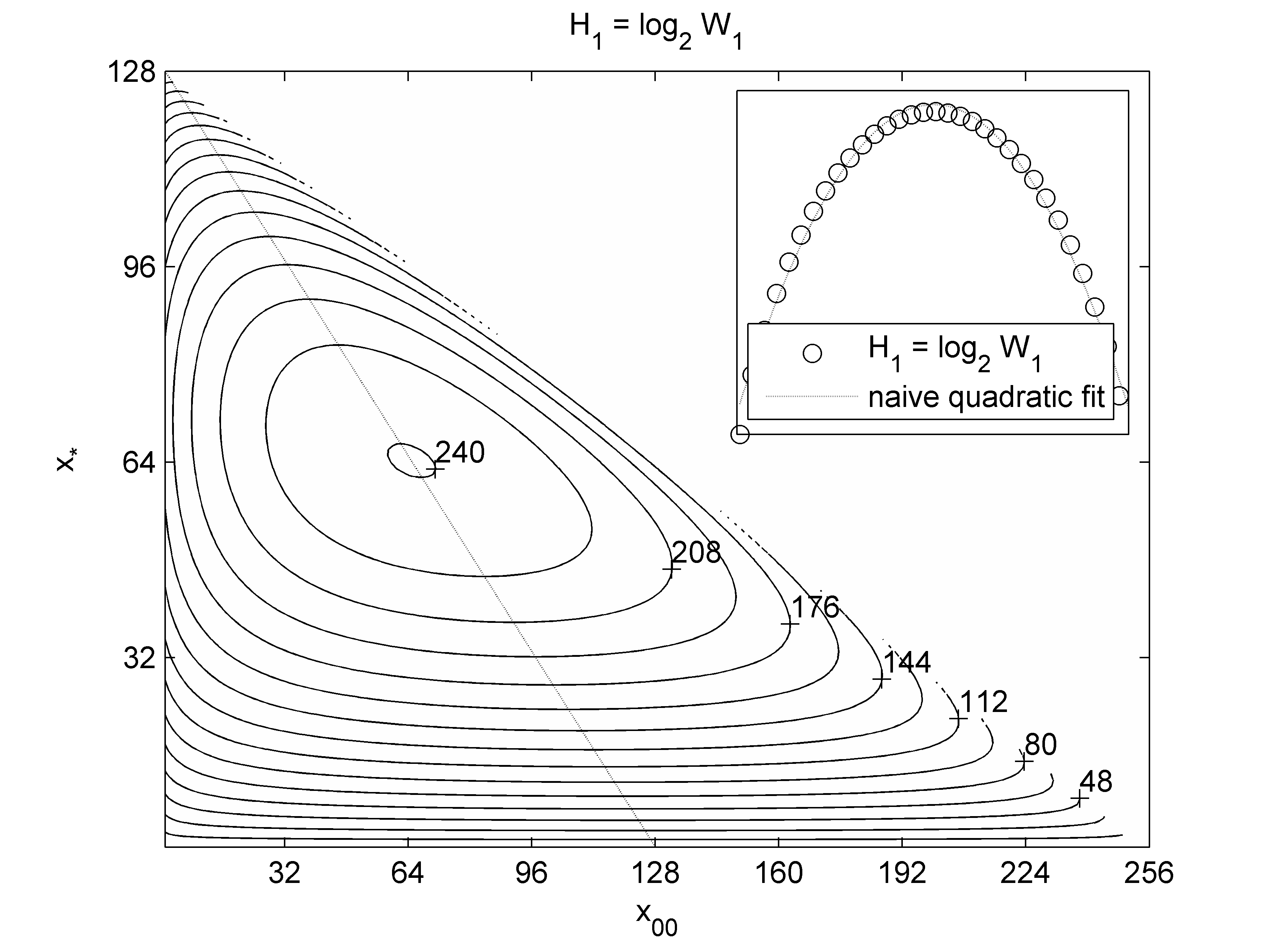}
\includegraphics[trim = 5mm 0mm 10mm 0mm, clip, width=80mm,keepaspectratio]{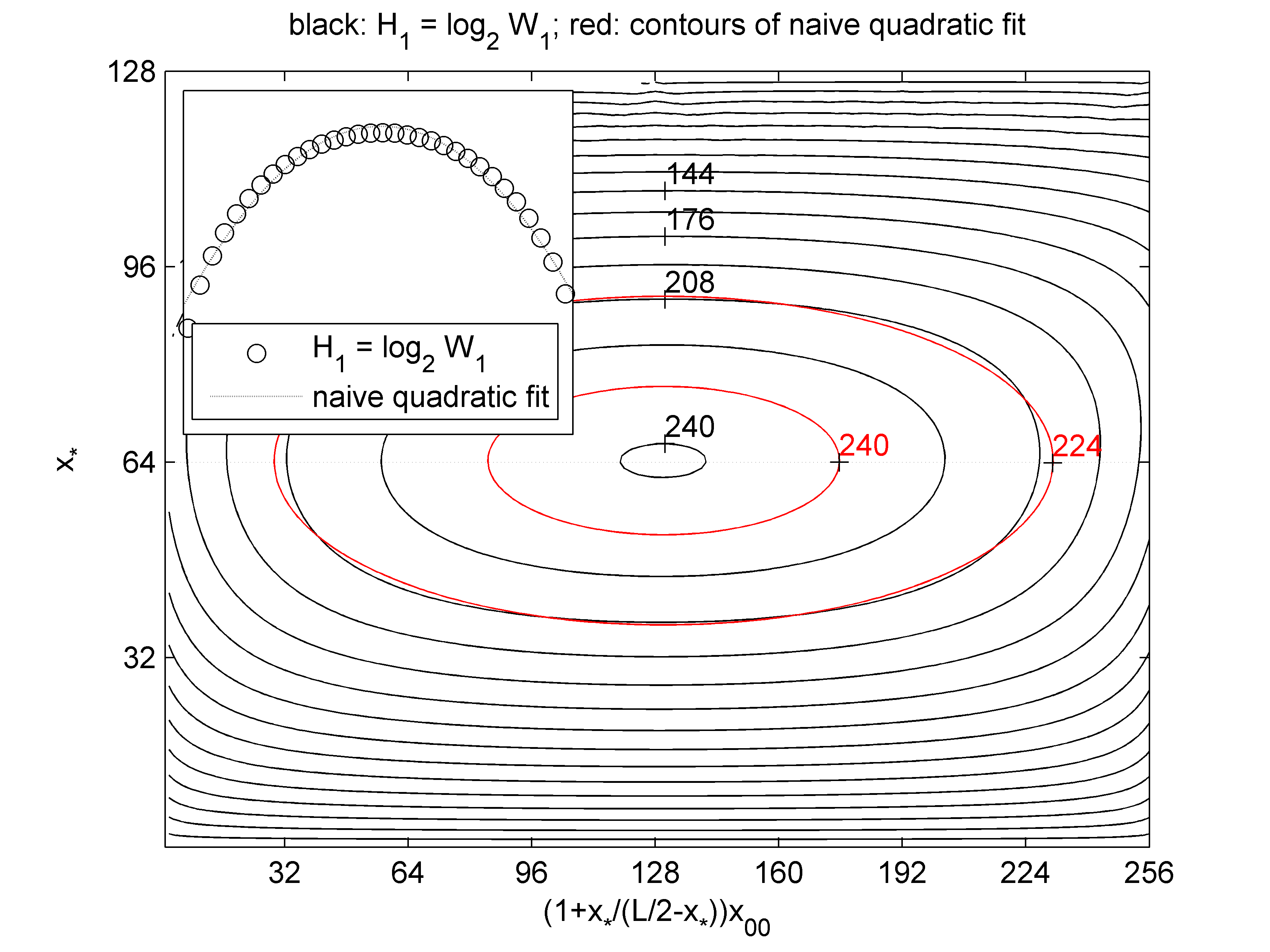}
\caption{ \label{fig:gaussian} (L) Contour plot of $H_1 := \log_2 W_1$ for $\ell = 256$. Inset: plot of $H_1$ values intermittently sampled along the dotted line and comparison to a naive quadratic fit. Samples along horizontal lines behave similarly. (R) Black countours: plot of $H_1$ after the horizontal transformation $x_{00} \mapsto (1 + \frac{x_*}{\ell/2-x_*}) \cdot x_{00}$. Red contours: naive quadratic fit. Inset: plot of transformed $H_1$ values intermittently sampled along the line $x_* = \ell/4$ and comparison to a naive quadratic fit. Note that the naive quadratic fit is a slight overestimate at the peak.} 
\end{figure}

Figure \ref{fig:gaussian} shows results in the same vein for $\ell = 256$. It is evident that $W_1$ behaves very much like a Gaussian, with the only significant qualitative difference resulting from the triangular domain. Similar results hold for more general contexts, and this fact might enable analytical estimates for $W(A)$. $\Box$

\section{\label{sec:InformationTheory}Information theory}

\subsection{\label{sec:deBruijnEntropy}de Bruijn entropy}

\textsc{Definition.} The \emph{order $k$ de Bruijn entropy} of a cyclic word $w$ is $H_k(w) := \log W_k(w)$. 

\

As in \S \ref{sec:Combinatorics}, we may also consider the entropy of an Eulerian quiver $Q$ or of its adjacency matrix $A$, written respectively $H(Q)$ and $H(A)$. Typically the logarithm will be taken with base $|\mathcal{A}|$ unless otherwise indicated. 

This definition evokes Boltzmann's physical interpretation of entropy as the logarithm of the number of microscopic configurations of a system that are consistent with the system's macroscopic characterization. Here, the ``macroscopic characterization'' of $w$ is just $Q_k(w)$, and ``microscopic configurations'' are just members of $[w]_k$. Another perspective realizes this definition as an analogue for finite words of the capacity of the discrete noiseless channel \emph{\`a la} Shannon \cite{Shannon}, or equivalently of the topological entropy of a subshift of finite type \cite{Kitchens}.

\subsubsection{\label{sec:CompressionArguments}Compression arguments}

Recall now the context of \S \ref{sec:example2}. In order to completely specify a cyclic binary word $w$, it suffices to specify both
\begin{itemize}
\item $A_1(w)$, which requires a total of $2 \lceil \log_2 \ell \rceil - 1$ bits (because it requires $\lceil \log_2 \ell \rceil$ bits to specify $x_{00}$ and $\lceil \log_2 \ell \rceil - 1$ bits to specify $x_*$);
\item The appropriate element of $[w]_1$, which requires at most $\lceil H_1(w) \rceil \lessapprox \ell - \log_2 \ell$ bits (because there are roughly $\ell^{-1} 2^\ell$ cyclic binary words of length $\ell$).
\end{itemize}
In particular, if $H_1(w) < \ell - 2 \log_2 \ell + 1$, then we have the outline of a scheme for losslessly compressing $w$ (the generalizations to $n > 2$ and $k > 1$ are not fundamentally different). Note that while most words are too statistically uniform (or more precisely, the adjacency matrices of their de Bruijn quivers have elements that are too similar) to be compressed in this way, in practice one is rarely interested in compressing statistically uniform data. Indeed, we recall that a standard diagonal argument shows that any fixed compression scheme will fail to compress most data \cite{LiVitanyi}.

The perspective of algorithmic information theory hinted at here will directly motivate the definition of relative de Bruijn entropy in \S \ref{sec:RelativeDeBruijnEntropy}.

\subsubsection{\label{sec:MaximallyInformativeValuesOfK}Maximally informative values of $k$}

Although the paper \cite{SimsEtAl} leverages the empirical probability distribution of $k$-tuples rather than the more detailed notion of de Bruijn quivers, it nevertheless gives strong experimental evidence that the natural heuristic $k = \lfloor \log_n \ell \rfloor$ is a good approximation for the lower limit of a reasonably narrow range of maximally informative values of $k$ in practice. While this paper also discusses upper limits on this range, these depend on the particular word and are of less practical interest for the obvious reason that increasing $k$ requires more storage.

\subsubsection{\label{sec:ComputationalComplexity}Remarks on computational complexity}

A detailed analysis of the complexity of computing the de Bruijn entropy is likely to be both more intricate and less informative than an experimental one, owing to the complex relationship between the local statistical behavior of words and their corresponding quiver connectivity structure as a function of $k$ (this is particularly true for the relative de Bruijn entropy, for which see below). However, we note that the dominant contribution to runtime is a matrix determinant (note that forming the adjacency matrices of quivers of words can be done in linear time), and we briefly discuss its complexity here.

Let $\omega$ denote the exponent for the complexity of matrix multiplication/inversion/determinant evaluation (say, 3 or perhaps $2.808$ in practice, or $2.373$ in theory \cite{VassilevskaWilliams}). Now $O((n^k)^\omega) = O(\ell) \iff k = \omega^{-1} \log_n \ell + O(1)$ determines $k$ s.t. computing the de Bruijn entropy requires linear time with standard techniques of linear algebra (e.g., computing the determinant via LU or QR decomposition as in our current implementation). Meanwhile, as pointed out in \S \ref{sec:MaximallyInformativeValuesOfK}, a reasonable rule of thumb for the maximally informative value of $k$ is $\lfloor \log_n \ell \rfloor$. 

These two observations can be combined by thinking of $k$ as a scale below which where we have complete information about the structure of words and of $\ell^{1/\omega}$ as a scale above which negligible information suitable for comparisons between words is discernable in linear time using standard techniques of linear algebra. That is, the computation of a de Bruijn entropy can be \emph{forced} to run in linear time by choosing $k = \lfloor \omega^{-1} \log_n \ell \rfloor$ (or, for that matter, $k = O(1)$), with the consequence that this amounts to neglecting not only correlations at scales greater than $k$ (as usual), but also the ability to capture statistical fluctuations \emph{of any sort} at scales beyond $\ell^{1/\omega}$. Insisting on $k = \lfloor \log_n \ell \rfloor$ means in practical terms that our technique requires cubic time in the implementation used here.

However, it is possible to do better, though for the sake of keeping this paper reasonably circumscribed we will confine ourselves here to a brief discussion. The reader will probably have noticed the phrase ``standard techniques of linear algebra'' repeated above, and considered the associated references to matrix decompositions for computing a determinant in (what is in practice) cubic time. In fact the determinant can be evaluated in less than $O((n^k)^\omega)$ time: it can be done in $O((n^k)^2)$ or even $O((n^k) \log^2 (n^k))$ time using so-called \emph{black box linear algebra} \cite{vonZurGathenGerhard,Wiedemann}. The key here is that a diagonal minor $\hat L$ of the Laplacian has a predetermined sparse structure, so that the oracle $x \mapsto \hat L x$ can be realized in subquadratic time. This faciliates the computation of the characteristic polynomial of $\hat L$ using so-called \emph{superfast} Toeplitz solvers in $O((n^k) \log^2 (n^k))$ time \cite{BrentEtAl,AmmarGragg}, from which the determinant follows trivially.
\footnote{
The notorious instability of superfast Toeplitz solvers for asymmetric matrices \cite{Bunch} is probably not a critical concern in the present context since the Laplacian matrix has integer entries; in any event, recently developed superfast solvers (see, e.g., \cite{XiaXiGu}) have also addressed this problem.
}

Thus although our current implementation essentially has cubic time complexity for maximally informative values of $k$, it can already be regarded as having linear complexity for $k$ independent of $\ell$, and a sufficiently optimized linear algebra subroutine would yield complexity that is just the product of a linear and a polylogarithmic term in the general regime of interest, rendering it competitive with bag-of-words or kernel methods \cite{Rieck} that have linear complexity but weaker or more \emph{ad hoc} theoretical justification.

\subsection{\label{sec:RelativeDeBruijnEntropy}Entropy of componentwise Eulerian quivers and relative de Bruijn entropy}

Write 
\begin{equation}
\label{eq:boxminus}
A \boxminus A' := (A-A') \vee 0 + (A'-A)^T \vee 0,
\end{equation}
where the maxima are taken elementwise. It is easy to see that if $A$ and $A'$ both correspond to componentwise Eulerian quivers, then so does $A \boxminus A'$. Indeed, this is the adjacency matrix of the quiver that naturally corresponds to $A-A'$ after reversing edges with negative matrix entries. \footnote{
Similarly, define $A \boxplus A' := A \boxminus A'^T$. If $A$ and $A'$ both correspond to Eulerian (resp., componentwise Eulerian) quivers, then so does $A \boxplus A'$. The operations $\boxplus$ and $\boxminus$ therefore induce an abelian semigroup structure on the set of (possibly degenerate) Eulerian quivers and an abelian group structure on the set of (possibly degenerate) componentwise Eulerian quivers.
}

With this in mind, let $A^{(j)}$ be adjacency matrices respectively corresponding to Eulerian quivers $Q^{(j)}$, so that $Q := \bigcup_j Q^{(j)}$ is a componentwise Eulerian quiver with corresponding adjacency matrix $A$. Define
\begin{equation}
\label{eq:WA}
W(A) := \prod_j W(A^{(j)})
\end{equation}
and
\begin{equation}
\label{eq:HA}
H(A) := \log W(A) = \sum_j H(A^{(j)}).
\end{equation}
To avoid degeneracies, we define $W(a) \equiv 1$ and $H(a) \equiv 0$, where here $a$ is the $1 \times 1$ adjacency matrix corresponding to the quiver $\Gamma_a$ with a single vertex and $a \ge 0$ edges (i.e., loops). This definition extends the prior one from Eulerian quivers to componentwise Eulerian quivers. Note that $H(A) = H(A^T)$ and $(A \boxminus A')^T = A' \boxminus A$, so that $H(A \boxminus A') = H(A' \boxminus A)$. 

Suppose now that we have two cyclic words $w$ and $w'$ over $\mathcal{A}$. Given $w$ and therefore also $A_k(w)$, all that is needed to determine $A_k(w')$ is the difference $A_k(w') - A_k(w)$, or equivalently the two nonnegative matrices 
\begin{eqnarray}
A_k(w|w') & := & [A_k(w)-A_k(w')] \vee 0; \\
A_k(w'|w) & := & [A_k(w')-A_k(w)] \vee 0.
\end{eqnarray} 
In order to completely specify $w'$ given $w$, it therefore suffices to specify $A_k(w|w')$, $A_k(w'|w)$, and a number of roughly $H_k(w')$ bits. It is clear that $H_{k+1}(w') \le H_k(w')$. If $w'$ is far from statistically uniform, then there will be some critical value $k << \ell(w')$ s.t. $H_k(w') = 0$ (note that $H_{\ell(w')-1}(w') \equiv 0$). At this point all the information in $w'$ that is not latent in $w$ is encoded in the matrices $A_k(w|w')$ and $A_k(w'|w)$. In other words, the conditional Kolmogorov complexity $K(w'|w)$ as well as the information distance \cite{BennettEtAl,LiEtAl2} can be approximated by a function of these matrices. 
\footnote{
We note in passing that the problem of comparing two words of vastly different length is sometimes of interest, and strategies such as those discussed in \cite{SimsEtAl} may be appropriate in certain (but certainly not all) contexts.
}

This motivates the following 

\

\textsc{Definition.} The \emph{order $k$ relative de Bruijn entropy} of $w'$ given $w$ is $H_k(w'||w) := H(A_k(w,w'))$, where $A_k(w,w') := A_k(w) \boxminus A_k(w') = A_k(w|w') + A_k^T(w'|w)$. More generally, the relative entropy of $A'$ given $A$ is defined as $H(A'||A) := H(A' \boxminus A)$. 

\

Note that (unlike the Kullback-Leiber incarnation of relative entropy for probability distributions) the relative entropy of componentwise Eulerian quivers is symmetric. Our experiments have shown that it is however not a pseudometric: i.e., it does not satisfy the triangle inequality. Nevertheless, it is straightforward to use the relative entropy to derive a pseudometric on a fixed set of words using the results of \cite{BrickellEtAl}.

\subsubsection{\label{sec:example3}Example 3}

Setting $u := ABRACADABRA$ and $v: = ABARACARBAD$, we have that $A_1(u) \boxminus A_1(v) = A_1(w)$, where $w := ABRABRABRA$ and $H_1(w) = 0$, so $H_1 (u || v) = 0$. 
\footnote{
Our MATLAB code readily computes this as \texttt{q1 = radixwordquiver('ABRACADABRA',1,'ABCDR'); q2 = radixwordquiver('ABARACARBAD',1,'ABCDR'); cequiverentropy(boxminus(q1,q2))}.
}
Meanwhile, the (Levenshtein) edit distance between $u$ and $v$ turns out to equal 5. It is evident at least in this particular case that the relative entropy better captures similarity in the local structure of words than an edit distance does.

\subsubsection{\label{sec:example4}Example 4}

For $1 < m \in \mathbb{N}$, let $w := 0^{m\ell} 1^{m\ell}$ and $w' := (0^\ell 1^\ell)^m$. For $k < \ell$, a straightforward (if somewhat tedious) calculation shows that $H_k(w || w') \lesssim m \log mk$, whereas the (Levenshtein) edit distance between $w$ and $w'$ is $2 \lfloor m/2 \rfloor \ell$. That is, for $k$ and $m$ fixed we have that $H_k(w || w') = O(1)$, whereas the corresponding edit distance is $O(\ell)$.

\subsubsection{\label{sec:example5}Example 5}

\begin{figure}[htbp]
\includegraphics[trim = 5mm 0mm 10mm 0mm, clip, width=80mm,keepaspectratio]{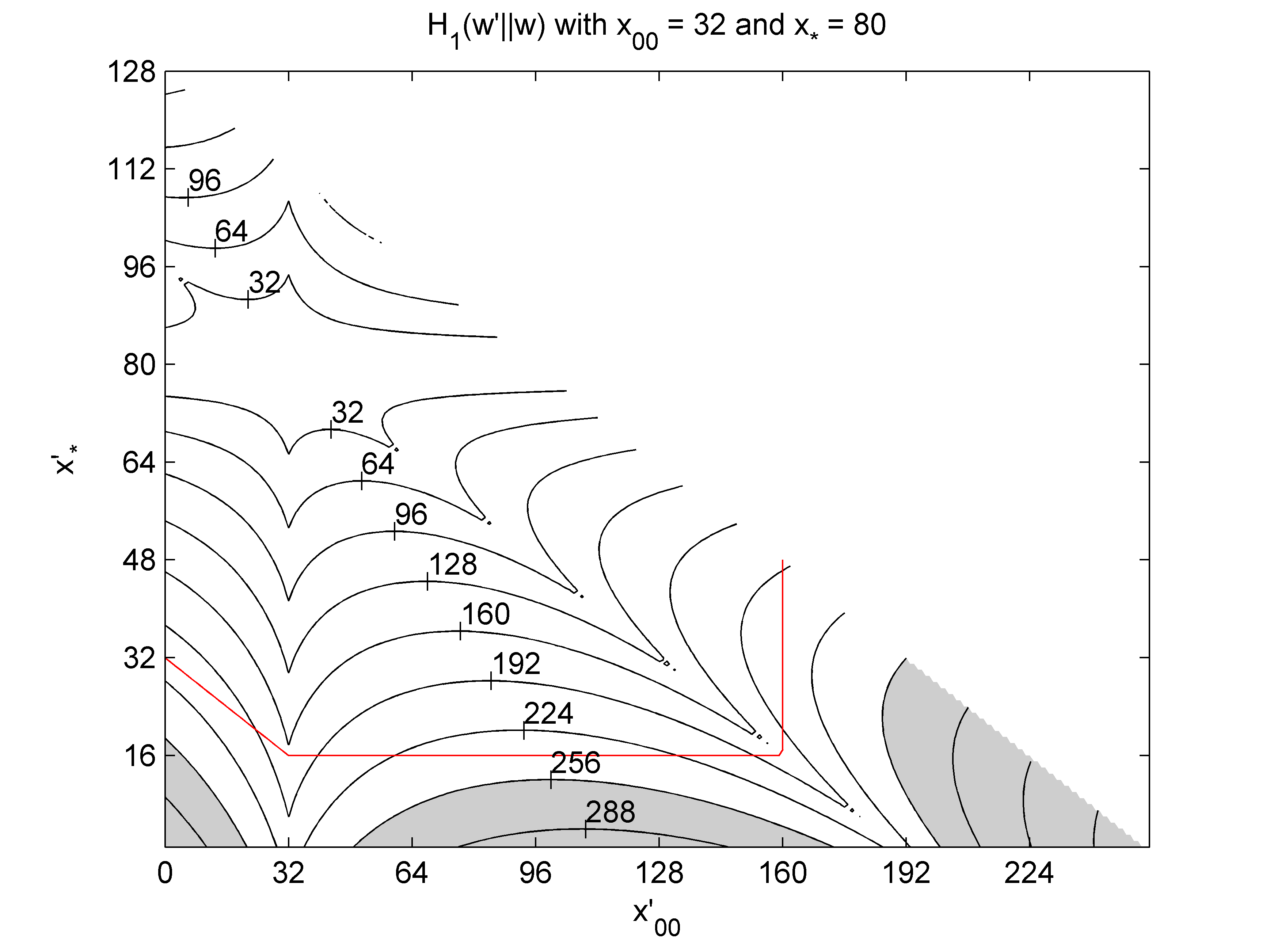}
\caption{ \label{fig:relative} Relative entropy $H_1(w'||w)$ of cyclic binary words $w, w'$ as a function of $x'_{00}$ and $x'_*$, where $\ell = 256$, $x_{00} = 32$, and $x_* = 80$. Regions where $H_1 > \ell$ are shaded (from left to right, they correspond to words close to $1 \dots 1$, $0 \dots 0 1 \dots 1$, and $0 \dots 0$). The red contour delineates where the sum of the entries of $A \boxminus A'$ equals $\ell$. This makes it clear that values of the relative entropy exceeding $\ell$ are due to a very large disparity between the input matrices. An alternative approach guaranteeing, e.g., $H(w'||w) < \ell(w) \vee \ell(w')$ would appear to either require working directly with the cycle spaces of $A$ and $A'$ or the blunt invocation of a threshold.} 
\end{figure}

Figure \ref{fig:relative} depicts the relative entropy $H_1(w'||w)$ of cyclic binary words $w, w'$ as a function of $x'_{00}$ and $x'_*$, where $\ell = 256$, $x_{00} = 32$, and $x_* = 80$.


	%

The relative entropy is zero along the strip $|x'_* - x_*| \le 1$ (and nowhere else). This is because in the strip, $A_1(w) \boxminus A_1(w')$ corresponds to coherently inserting and/or deleting only cyclic subwords of the form $w_{(1)} = 0 \dots 0$, $w_{(2)} = 01 $, and $w_{(3)} = 1 \dots 1$. While there are generally many ways to do this, the cyclic subword $w_{(1)}w_{(2)}w_{(3)} = 0 \dots 01 \dots 1$ satisfies $H_1(w_{(1)}w_{(2)}w_{(3)}) = 0$. This in turn is a manifestation of the simple cycle structure of $A_1(w) \boxminus A_1(w')$ in the strip.

The cycle structure of $A_1(w) \boxminus A_1(w')$ is also behind the more significant phenomenon of relative entropy values exceeding $\ell(w) \vee \ell(w')$. This highlights the constraint that any modified relative de Bruijn entropies of the form $H(f(A_k(w),A_k(w')))$ should be such that $f(A_k(w),A_k(w'))$ is still a componentwise Eulerian adjacency matrix.



\section{\label{sec:Applications}Applications}

\subsection{\label{sec:Systematics}Biological systematics}

Molecular phylogenetics--i.e., the analysis of evolutionary relationships based on hereditary molecular characteristics--and biological classification of organisms typically focus on comparing DNA sequences \cite{ChanRagan}. A particularly convenient form of DNA for this purposes is mitochondrial DNA (mtDNA). mtDNA is an extremely economical repository of information in that nearly every base pair in human mtDNA is known to code for some protein or RNA product, and there is even overlap between coding regions; meanwhile mammalian mtDNA sequences are only on the order of 20k base pairs. Moreover, mtDNA is not highly conserved and mutates rapidly.

\begin{figure}[htbp]
\includegraphics[trim = 0mm 0mm 0mm 0mm, clip, width=140mm,keepaspectratio]{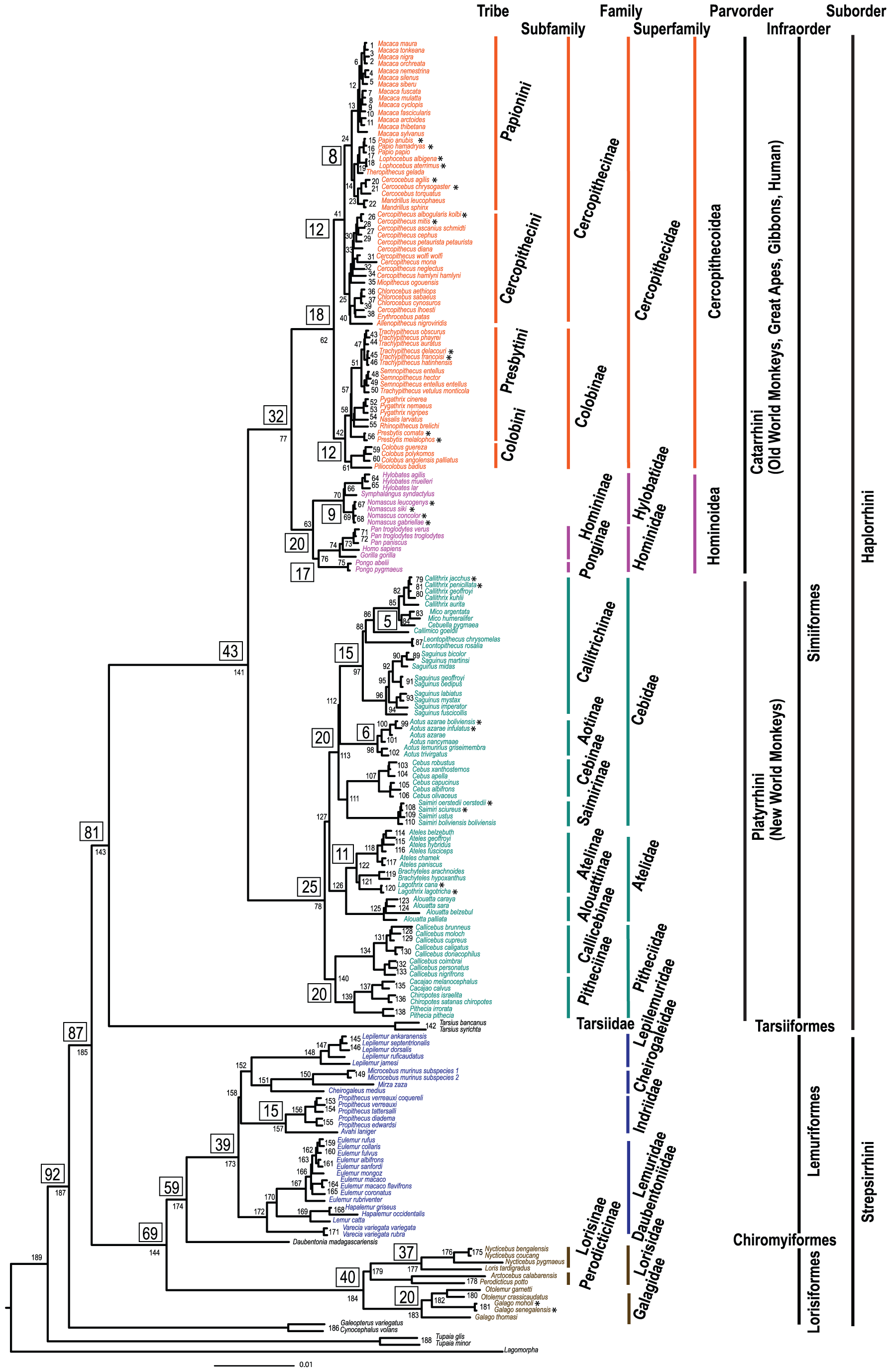}
\caption{ \label{fig:primatetree} ``The molecular phylogeny of 186 primates and four species representing the two outgroup orders of \emph{Scandentia}, \emph{Dermoptera}, and rooted by \emph{Lagomorpha}''. Figure and quotation taken from \cite{PerelmanEtAl} and used under the CC-BY-2.5 license (see http://creativecommons.org/licenses/by/2.5/).} 
\end{figure}

\begin{figure}[htbp]
\includegraphics[trim = 0mm 40mm 0mm 30mm, clip, width=160mm,keepaspectratio]{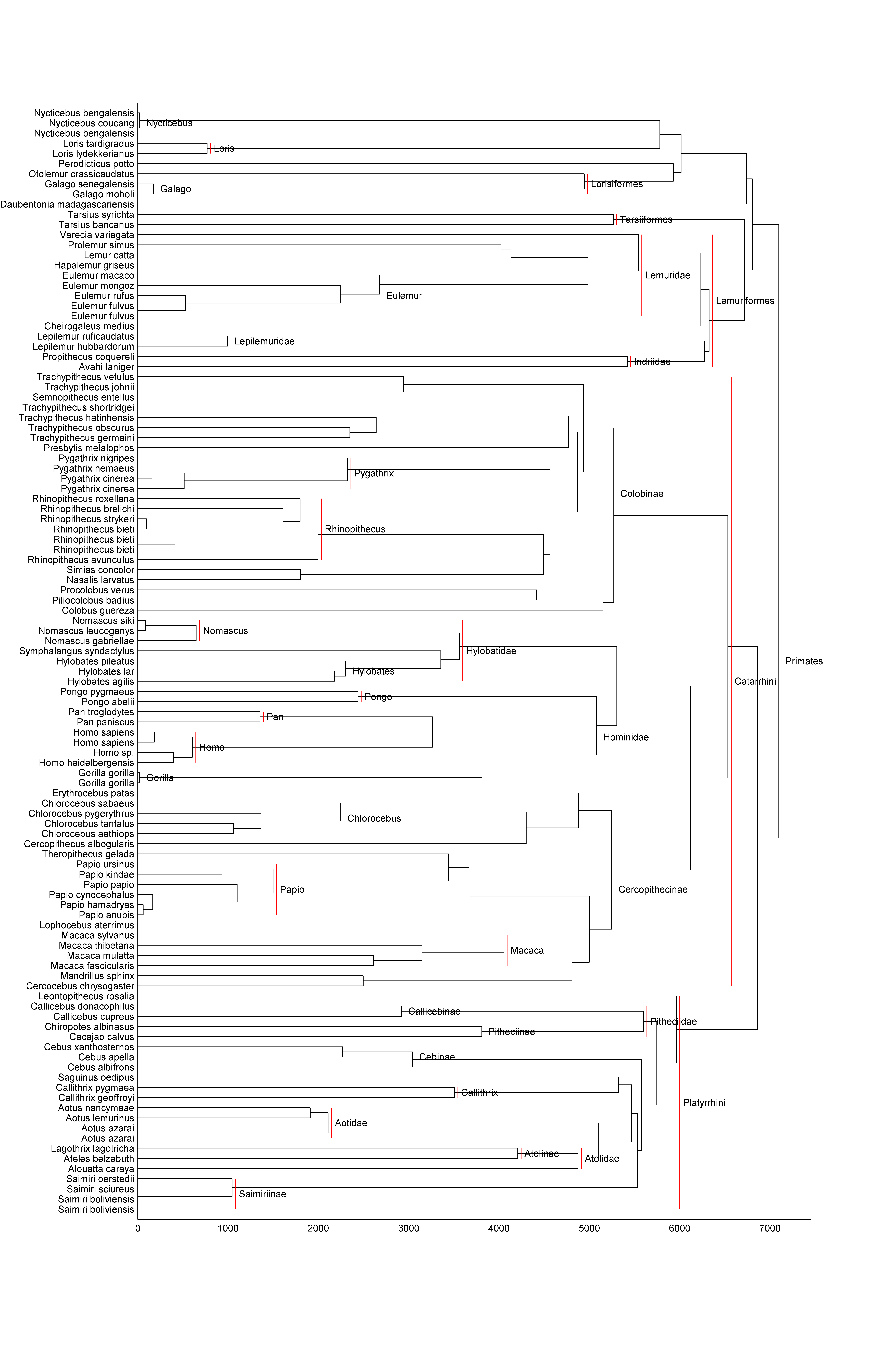}
\caption{ \label{fig:k7avg} Automatically generated (cf. \S \ref{sec:DataFigureGenSystematics}) phylogenetic tree using average linkage for $k = 7$ relative de Bruijn entropy (unnormalized). Mismatches w/r/t the tree in figure \ref{fig:primatetree} appear for the suborders \emph{Haplorrhini} and \emph{Strepsirrhini}, for the families \emph{Cebidae} and \emph{Lorisidae}, and for the subfamily \emph{Callitrichinae}. Note that \emph{Lorisiformes} here should be labeled as \emph{Galagidae}, but this merely reflects an ambiguity in the input data annotation. Not explicitly shown, but also matched, are the tribes \emph{Cercopithecini}, \emph{Colobini}, \emph{Papionini}, and \emph{Presbytini}, the subfamilies \emph{Homininae} and \emph{Lorisnae}, the superfamily \emph{Hominoidea}, and the infraorders \emph{Lorisformes} (cf. previous comment) and \emph{Simiformes}.} 
\end{figure}

\begin{figure}[htbp]
\includegraphics[trim = 0mm 40mm 0mm 30mm, clip, width=160mm,keepaspectratio]{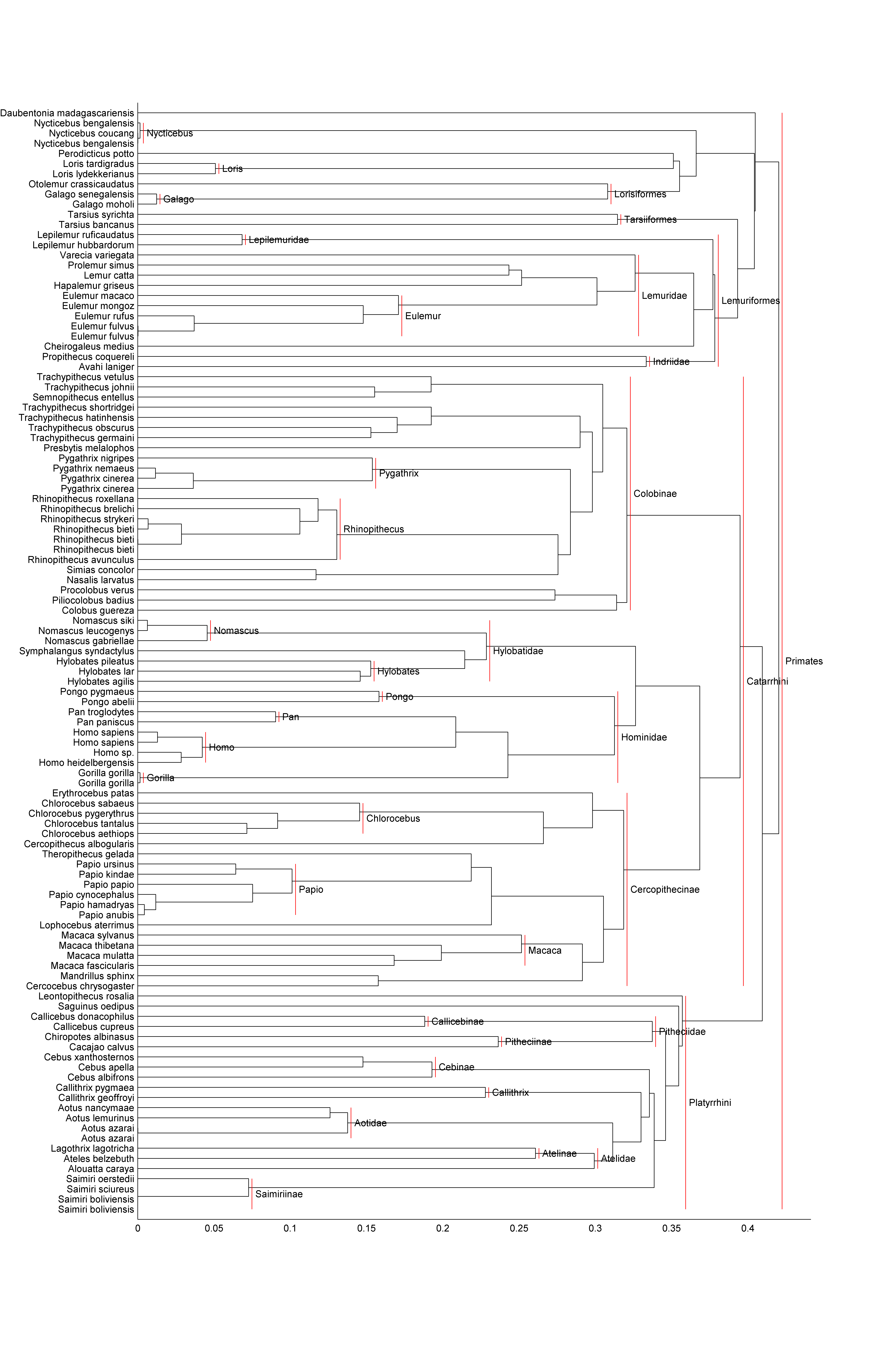}
\caption{ \label{fig:k7avgnorm} Phylogenetic tree using average linkage for $k = 7$ relative de Bruijn entropy (normalized). Note that this and the previous figure are quantitatively very similar. The mismatches are the same as in the previous figure, with the addition of \emph{Lorisinae}; apart from this, matched but unshown taxnomical groupings are also as in the previous figure.} 
\end{figure}

\begin{figure}[htbp]
\includegraphics[trim = 0mm 40mm 0mm 30mm, clip, width=160mm,keepaspectratio]{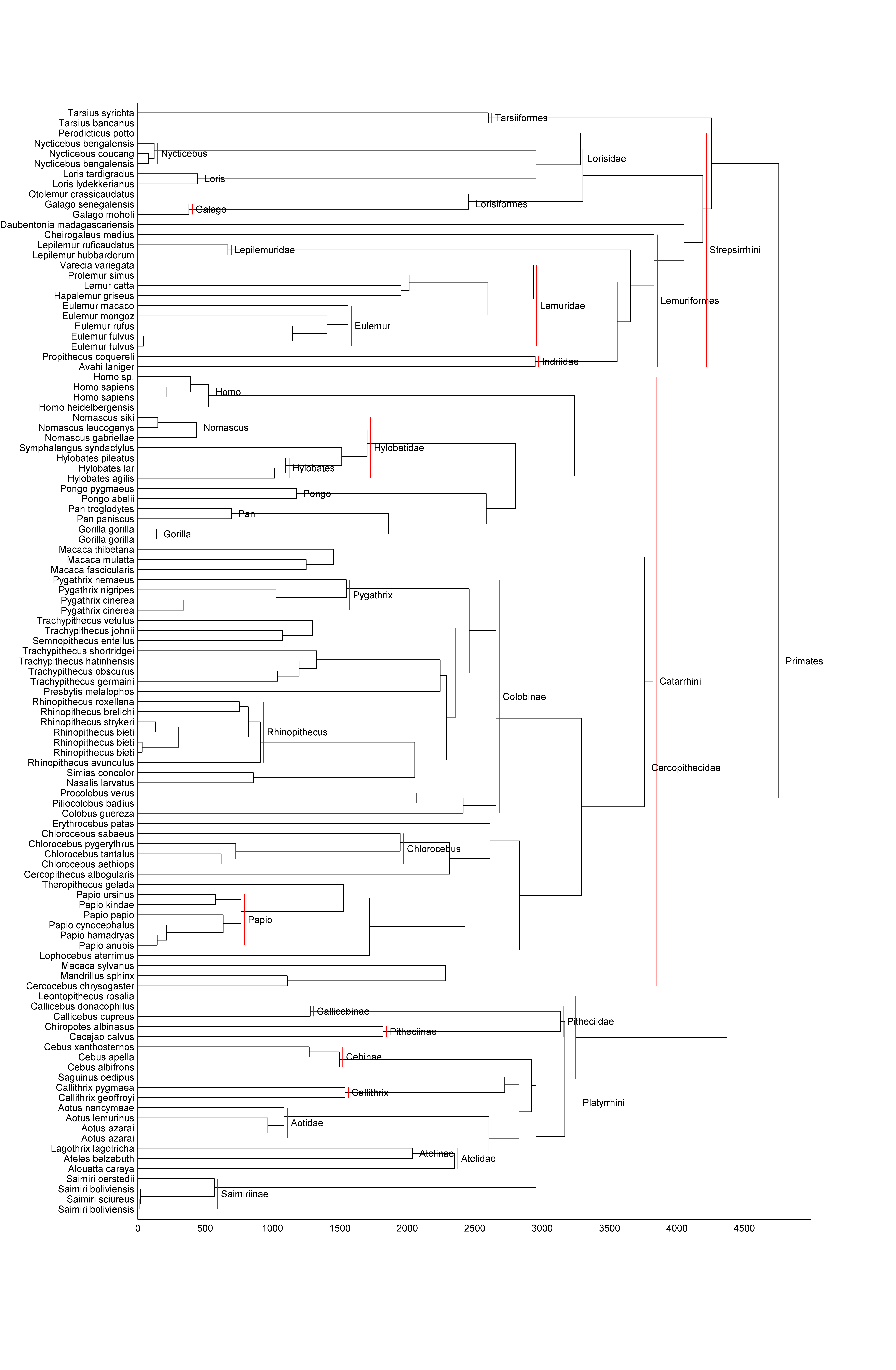}
\caption{ \label{fig:ldavg} Average linkage for (Levenshtein) edit distance (unnormalized). Note that while the edit distance results match \emph{Strepsirrhini} and \emph{Lorisidae}, they fail to match the more fine-grained taxa \emph{Macaca} and \emph{Homindae}, the latter of which is particularly important to members of, e.g., \emph{Homo sapiens} and is detailed in figure \ref{fig:comparison}. (NB. \emph{Cercopithecinae} and \emph{Cercopithecidae} are ambiguously labeled here and in previous figures due to the input data annotation and as such are not remarked on for comparative evaluation.)} 
\end{figure}

\begin{figure}[htbp]
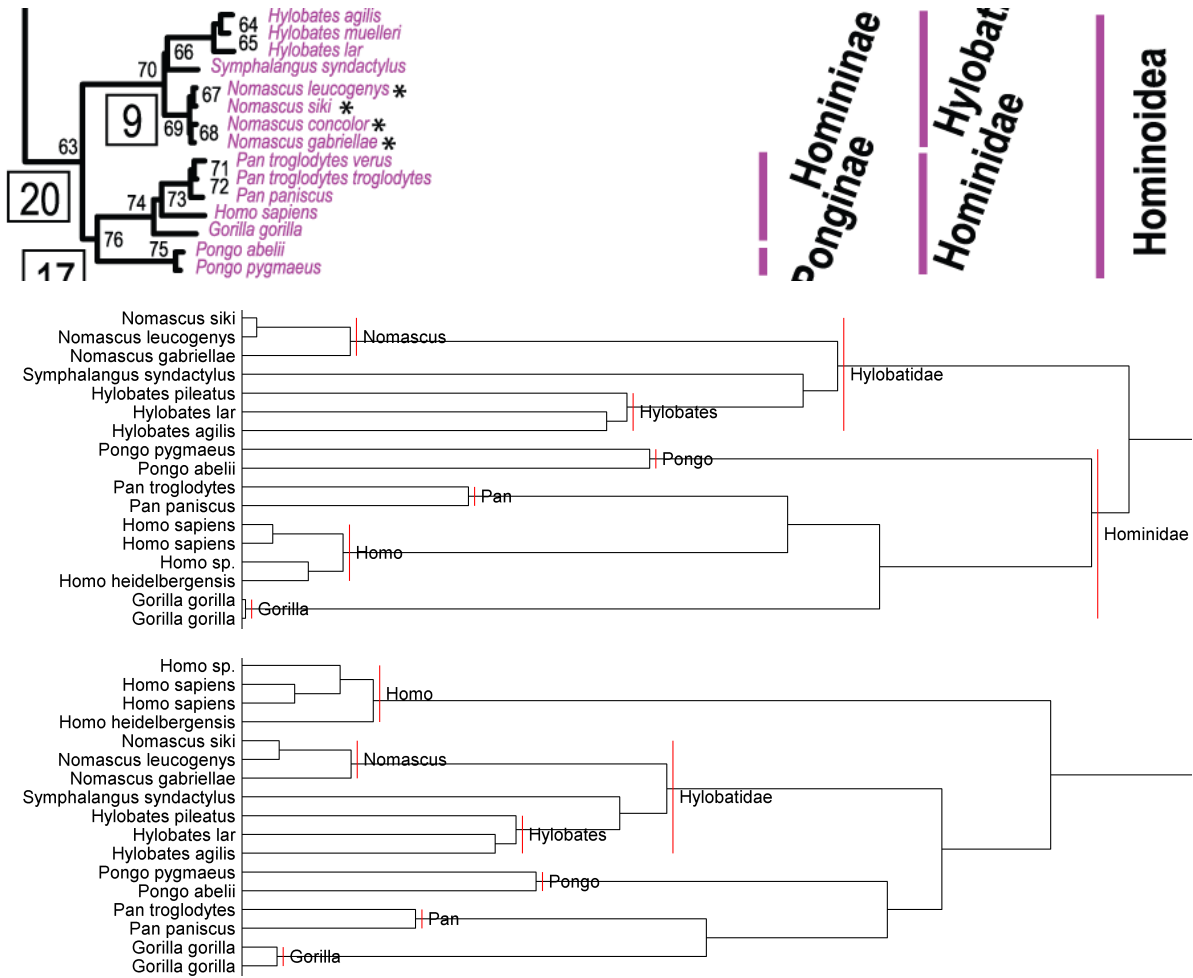

\includegraphics[trim = 120mm 345mm 70mm 204mm, clip, width=160mm,keepaspectratio]{primatetree.png}

\

\includegraphics[trim = 0mm 170.5mm 72mm 180.75mm, clip, width=160mm,keepaspectratio]{k7avg.png}

\

\includegraphics[trim = 0mm 242mm 72mm 109.25mm, clip, width=160mm,keepaspectratio]{ldavg.png}
\caption{ \label{fig:comparison} Detail of \emph{Hominoidea} from figures \ref{fig:primatetree}, \ref{fig:k7avg}, and \ref{fig:ldavg}, respectively. The local restrictions of trees from figures \ref{fig:primatetree} and \ref{fig:k7avg} to common species are topologically equivalent (note that the trees have internal symmetries), whereas the local restriction of the tree from \ref{fig:ldavg} disagrees with them.
} 
\end{figure}

In figures \ref{fig:k7avg} and \ref{fig:k7avgnorm} below we show that relative de Bruijn entropy produces results comparable if not superior to an edit distance (cf. figure \ref{fig:ldavg}) for constructing phylogenetic trees that easily capture most of the evolutionary relationships among primates (cf. figures \ref{fig:primatetree} and \ref{fig:comparison}) from mtDNA sequences alone. Furthermore, the comparative performance of the relative entropy is likely to improve in other problem domains that can actually leverage the fact that the relative entropy captures local correlations while ignoring global correlations (cf. \S \ref{sec:VET}). 

It is worth noting that the technique outlined here is \emph{alignment-free} \cite{ChanRagan}, and a suitable implementation optimized for speed (which our current implementation certainly is not, cf. \S \ref{sec:ComputationalComplexity}) is a promising candidate tool for bioinformatics. In particular, it is an attractive alternative to current techniques such as those in \cite{LiXuHao,WangEtAl,SimsEtAl,SimsKim,HatjeKollmar} and the older but perhaps conceptually closer approach of \cite{LiEtAl}. We note also that de Bruijn quivers have been considered in the context of multiple alignment \cite{RaphaelEtAl,ZhangWaterman,ZhangWaterman2}.

\subsection{\label{sec:Textual}Textual comparison}

There are numerous downstream applications of textual comparison, e.g. authorship attribution (see, e.g., \cite{SegarraEisenRibeiro}). 
%
%
It may be fruitful to engage in a dramatic simplification of textual comparison at large scales with relative de Bruijn entropy by, e.g., working with the results of part-of-speech tagging (perhaps focusing in particular on function words versus content words). Another possibility is to omit or ``don't care'' words that do not frequently occur in a corpus and treat the words themselves as symbols in a very large alphabet while keeping the order of the induced quivers very small.

\subsection{\label{sec:VET}Behavioral analysis of computer programs}

Notwithstanding the interesting applications discussed above, the \emph{raison d'\^{e}tre} of the present paper is the anticipated application of its theory to the low-level behavioral analysis of computer programs of unknown provenance. Although the general problem of disassembly of binary executable code is undecidable, interactive disassembly and automated reverse engineering techniques can facilitate static code analysis. In particular, a disassembled binary executable program may be conveniently represented in a platform-independent intermediate language such as REIL \cite{DullienPorst} or an enhanced variant called Power-REIL (PREIL). In turn, individual machine-level instructions (or short sequences of them) can be mapped to a reduced set of behaviors, which serve as the symbols in a prospective alphabet. One or more of these behaviors or symbols may serve as a flavor of ``don't care''.

An abstract view of the preceding enterprise consists of annotating (a suitable coarse-graining of) the program's control flow graph with the appropriate behaviors: vertices are annotated with a control flow behavior, and edges are annotated with a sequence of other behaviors. Specific executions of the program correspond to walks in the control flow graph, which induce corresponding sequences of behaviors--i.e., words over the behavior alphabet. Broad coverage of the control flow graph can be accelerated through dynamic execution \cite{KoushikMarinovAgha}. At this point one has a set of words corresponding to representative program behavior sequences. Comparing these (or extracted subwords) to words containing exemplars of bad behavior can then inform security assessments \cite{CesareXiang}.

The rationale for using relative de Bruijn entropy here versus a more familar construct such as edit distance is straightforward. The relative de Bruijn entropy compares all of the local structure of words while remaining agnostic to the global structure. Because of this, it is particularly well suited to comparing words corresponding to different paths through a control flow graph, where ``motifs'' may appear with greater location variability than in biological sequences. \emph{In particular, it is natural to expect that loops containing complex sub-flow graphs are better suited to analysis with relative de Bruijn entropy than an edit or dynamic programming metric on words.}

\acknowledgements

The authors are grateful to 
R. Ross for his helpful comments and assistance. Prior conversations between the first author and H. Fredricksen and D. H. Wood also substantively informed the discussion here. We are also grateful to referees of WORDS 2015, where a brief version of this paper was presented.
{\color{red}
}

\appendix

\section{\label{sec:Spin}Spin models}

In this appendix, we briefly mention the simple connection (via the transfer matrix method) of de Bruijn quivers and entropies with the physics of finite one-dimensional spin models.

A \emph{local potential} of order $k$ is a function $\mathcal{E} : \mathcal{A}^k \rightarrow \mathbb{R}$. The corresponding \emph{energy} of a cyclic word $w$ with $\ell(w) \ge k$ is 
\begin{equation}
\label{eq:energy}
\mathcal{E}(w) := \sum_{j=1}^{\ell(w)} \mathcal{E}(w_j \dots w_{j+k-1}).
\end{equation}
A probability distribution of the form $\mathbb{P}(w) = e^{-\beta \mathcal{E}(w)}/Z(\beta)$ over cyclic words of some fixed length $\ell$ defines the \emph{canonical ensemble} of a circular one-dimensional \emph{spin model}, or more generally a Gibbs field \cite{Bremaud}. In statistical physics, $\beta = 1/k_B T$, where $k_B$ is the Boltzmann constant, $T$ is the absolute temperature, and the normalizing factor $Z(\beta)$ is the so-called \emph{partition function}. It turns out that every quantity of physical interest (e.g., entropy, internal energy, free energy, etc.) can be computed from a system's partition function, and its determination is correspondingly the central goal of statistical physics \cite{MezardMontanari}.

For example, taking $\mathcal{A} = \{0,1\}$, $\sigma_j := 2w_j-1$, and $\mathcal{E}(w_j w_{j+1}) := -2J \sigma_j \sigma_{j+1} - K \sigma_j$ reproduces the 1D (spin-1/2) Ising model describing model magnetic systems. Here the spins $\sigma_j$ represent magnetic dipoles, $J$ represents the strength of the dipole-dipole coupling, and $K$ represents the external magnetic field. Another example is furnished by taking $\mathcal{A} = \{A,C,G,T\}$ and defining a local potential by the nearest-neighbor Gibbs free energy parameters quantitatively describing oligomeric DNA hybridization \cite{SantaLucia}. 
\footnote{
As an aside, the physics of DNA indicate that this local energy reflects vibrational degrees of freedom, and in principle similar local energies on $\mathcal{A}^3$ and $\mathcal{A}^4$ could reflect bending and torsional degrees of freedom. However, since the bending and torsional modes are so much more easily excited than the vibrational modes, there is unlikely to be any significant improvement from addressing such considerations. 
}

The transfer matrix method facilitates the calculation of $Z^{1/\ell}$ and its limit as $\ell \rightarrow \infty$. For example, it is a standard result that for the 1D Ising model
\begin{equation}
\label{eq:isinglimit}
\lim_{\ell \rightarrow \infty} Z^{1/\ell} = e^{\beta J} \cosh \beta K + \sqrt{e^{2 \beta J} \sinh^2 \beta K + e^{-2 \beta J}}.
\end{equation}
The particular flavor of transfer matrix method embodied in \eqref{eq:example2eq} enables the exact calculation of $Z^{1/\ell}$ for $\ell < \infty$ by noting that
\begin{equation}
\label{eq:densityofstates}
Z = \sum_w e^{-\beta \mathcal{E}(w)} = \sum_{x_{00},x_*} W_1 e^{-\beta \mathcal{E}} = \sum_{x_{00},x_*} e^{H_1-\beta \mathcal{E}}
\end{equation}
along with
\begin{equation}
\label{eq:isingenergy}
\mathcal{E}(w) \equiv \mathcal{E}(x_{00},x_*;\ell) = 2Kx_{00} + (4J+2K)x_* - (J+K)\ell.
\end{equation}
This is superfically rather different than writing an expression of the form $Z = \text{Tr}(\Lambda^\ell)$ as per a typical application of the transfer matrix method in physics, but in fact both approaches turn out to have the same essential content and structure \cite{Stanley}.

%

\section{\label{sec:DataFigureGen}Data and figure generation}

NB. MATLAB code detailed in \S \ref{sec:MATLABcode} is called throughout this section.

\subsection{\label{sec:DataFigureGenExample2} \S \ref{sec:example2}}

The figures in \S \ref{sec:example2} were generated using the following MATLAB commands:

\footnotesize
\begin{verbatim}
A1 = 4*[2,5;5,4];
L = sum(sum(A1));	% 256
e = nan(L/2,L);
ce = nan(L/2,L);
e2 = nan(L/2,L);
ce2 = nan(L/2,L);
for j = 1:L/2
   for k = 1:(L-1)
       [j k] % for tracking progress
       A2 = [k-1,j;j,L-(k-1)-2*j]; % same length
       Aa = max(A1-A2,0);
       Ab = max(A2-A1,0);
       A = Aa'+Ab;
       if any(any(A2<0)) % then A2 is ill-formed
       else
           e(j,k) = equiverentropy(A);
           ce(j,k) = cequiverentropy(A);
       end
       e2(j,k) = equiverentropy(A2);
       ce2(j,k) = cequiverentropy(A2);
   end
end

H = ce2/log(2);
figure;[C,h] = contour(H,16:16:256,'k');
clabel(C,'manual');
hold;plot((1:L/2),L/2-(1:L/2)+1,'k:');
set(gca,'XTick',0:32:256)
set(gca,'YTick',0:32:128)
xlabel('x_0_0');
ylabel('x_*');
title('H_1 = log_2 W_1');

Hmid = mean(H(:,[L/2,L/2+1]),2);
ind = find(~isnan(Hmid));
p = polyfit(ind,Hmid(ind),2);
figure;plot(1:4:(L/2),Hmid(1:4:end),'ko',1:(L/2),p(1)*(1:(L/2)).^2+p(2)*(1:(L/2))+p(3),'k:');
axis([0 L/2 0 L]);
set(gca,'XTick',[]);
set(gca,'YTick',[]);
xlabel({'x_* with (x_0_0,x_*) intermittently sampled','along vertical domain bisector'});
h = legend('H_1 = log_2 W_1','naive quadratic fit',4);
pvert = p;

Hmid = mean(H([L/4,L/4+1],:),1);
ind = find(~isnan(Hmid));
p = polyfit(ind,Hmid(ind),2);
figure;plot(ind(1:8:end),Hmid(ind(1:8:end)),'ko',1:L,p(1)*(1:L).^2+p(2)*(1:L)+p(3),'k:');
axis([0 L L/2 L]);
set(gca,'XTick',[]);
set(gca,'YTick',[]);
xlabel({'x_* with (x_0_0,x_*) intermittently sampled','along horizontal domain bisector'});
h = legend('H_1 = log_2 W_1','naive quadratic fit',4);
phorz = p;

[x,y] = meshgrid(1:L,1:L/2);
x2 = x+(x.*y)./(L/2-y); y2 = y;
ind = intersect(find(~isnan(H)),find(~isinf(x2)));
w = griddata(x2(ind),y2(ind),H(ind),x,y);

figure;[C,h] = contour(w,16:16:L,'k'); 
clabel(C,'manual');
hold;plot(1:L,L/4,'k--');
g = phorz(1)*x.^2+phorz(2)*x+phorz(3)+pvert(1)*y.^2+pvert(2)*y+pvert(3);
g = g/2;    % horz and vert double count
[C2,h2] = contour(g,[224,240],'r');
clabel(C2,'manual','Color','r');
set(gca,'XTick',0:32:256)
set(gca,'YTick',0:32:128)
xlabel('(1+x_*/(L/2-x_*))x_0_0');
ylabel('x_*');
title('black: H_1 = log_2 W_1; red: contours of naive quadratic fit');
\end{verbatim}
\normalsize
\vskip\medskipamount 
\leaders\vrule width \textwidth\vskip0.4pt 
\vskip\medskipamount 
\nointerlineskip

\subsection{\label{sec:DataFigureGenexample5} \S \ref{sec:example5}}

Figure \ref{fig:relative} in \S \ref{sec:example5} was generated using the following MATLAB commands:

\footnotesize
\begin{verbatim}
A1 = 4*[2,5;5,4];
L = sum(sum(A1)); % 256
ce = nan(L/2,L);
ce2 = nan(L/2,L);
for j = 1:L/2
   for k = 1:(L-1)
       [j k] % for tracking progress
       A2 = [k-1,j;j,L-(k-1)-2*j]; % same length
       Aa = max(A1-A2,0);
       Ab = max(A2-A1,0);
       A = Aa'+Ab;
       if any(any(A2<0))
       else
           ce(j,k) = cequiverentropy(A);
       end
       ce2(j,k) = cequiverentropy(A2);
   end
end

LL = nan(L/2,L);
LM = nan(L/2,L);
for j = 1:L/2
   for k = 1:(L-1)
       A2 = [k-1,j;j,L-(k-1)-2*j]; % same length
       Aa = max(A1-A2,0);
       Ab = max(A2-A1,0);
       A = Aa'+Ab;
       if any(any(A2<0))
       else
           LL(j,k) = sum(sum(A));
           LM(j,k) = 1;
       end
   end
end

delta = L/8;
figure; 
contourf(ce/log(2),[L,L],'k','LineColor','none');
colormap(gray);
hold;
[C,h] = contour(ce/log(2),delta:delta:2*L,'k');
contour(LL,[L,L],'r')
caxis([0 L*1.25]);
set(gca,'XTick',1:delta:(L+1),'XTickLabel',0:delta:L)
set(gca,'YTick',0:delta/2:L/2)
xlabel('x''_0_0');
ylabel('x''_*');
title('H_1(w''||w) with x_0_0 = 32 and x_* = 80');
clabel(C,'manual')
\end{verbatim}
\normalsize
\vskip\medskipamount 
\leaders\vrule width \textwidth\vskip0.4pt 
\vskip\medskipamount 
\nointerlineskip

\subsection{\label{sec:DataFigureGenSystematics} \S \ref{sec:Systematics}}

The following commands were applied to two input text files: \texttt{genbankfile} and \texttt{fastafile}, which respectively contained 109 GenBank and FASTA-formatted mtDNA sequences of a broadly representative set of primates, and obtained as described in the code comments in \S \ref{sec:fastaparse}. \texttt{genbankfile} was used purely for its annotations; the remainder of the analysis used \texttt{fastafile}. The computation required less than 2 hours in a single MATLAB session on a standard laptop and automatically generated figures \ref{fig:k7avg} and \ref{fig:k7avgnorm}. Figure \ref{fig:ldavg} was generated along similar lines using data produced by the second author from a Levenshtein distance routine in Python that is not included here.

\footnotesize
\begin{verbatim}
%% Preliminaries
fp = fastaparse(fastafile);
N = length(fp.seq);
kmax = 7;
alpha = 'ACGTN';
%% Radix word quivers
for k = 1:kmax
    for j = 1:N
        rwq{k}{j} = radixwordquiver(fp.seq{j},k,alpha); 
        [j,k] % for tracking progress
    end 
end
disp('radix word quivers done');
%% Entropies of individual words
for k = 1:kmax
    H1{k} = zeros(1,N); 
    for j = 1:N
        H1{k}(j) = cequiverentropy(rwq{k}{j});
    end
    k % for tracking progress
end
disp('individual entropies done');
%% Relative entropies
for k = 1:kmax
    H{k} = zeros(N); 
    for ja = 1:N
        for jb = ja:N
            M = boxminus(rwq{k}{ja},rwq{k}{jb});
            H{k}(ja,jb) = cequiverentropy(M);
        end
        [ja,k] % for tracking progress
    end
    H{k} = H{k}+H{k}';
end
disp('relative entropies done');
%% Construct radix quivers for concatenated words and compute their entropy
% This is significantly more efficient than forming the radix quivers 
% directly
n = length(alpha);
for k = 1:kmax
    for i = 1:N
        u = fp.seq{i};
        Au = rwq{k}{i};
        for j = i:N
            v = fp.seq{j};
            Av = rwq{k}{j};
            crwq = concatenateradixwordquiver(u,v,Au,Av,alpha);
            Hcrwq{k}(i,j) = cequiverentropy(crwq);
        end
        [i,k] % for tracking progress
    end
    % Symmetrize and avoid double-counting the diagonal
    Hcrwq{k} = Hcrwq{k}+Hcrwq{k}'-diag(diag(Hcrwq{k}));
end
disp('concatenated radix quivers and entropies done');
%% Linkages
for k = 1:kmax
    sf{k} = squareform(H{k});
    sfnorm{k} = squareform(H{k}./Hcrwq{k});
    tree{k} = linkage(sf{k});
    treeav{k} = linkage(sf{k},'average');
    treecp{k} = linkage(sf{k},'complete');
    treenorm{k} = linkage(sfnorm{k});
    treenormav{k} = linkage(sfnorm{k},'average');
    treenormcp{k} = linkage(sfnorm{k},'complete');
    k % for tracking progress
end
disp('linkages done');
%% Taxonomic info
% genus/species
for j = 1:N
    i1 = max(strfind(fp.desc{j},'|'))+2;
    i2 = min(strfind(fp.desc{j},'mitochondrion, complete genome'))-1;
    desc{j} = fp.desc{j}(i1:i2); 
    [genus{j},remain] = strtok(desc{j},' ');
    species{j} = strtok(remain,' '); 
    genusspecies{j} = [genus{j},' ',species{j}];
end
% Other info
taxa = genbankorganism(genbankfile);
% Exclude taxonomic info above order (for Primates)
for j = 1:N
    taxa{j} = taxa{j}(10:end);
end
%% Produce phylogenetic trees
phylotree(treeav{kmax},genusspecies,taxa);
phylotree(treenormav{kmax},genusspecies,taxa);
phylotree(treenormcp{kmax},genusspecies,taxa);
\end{verbatim}
\normalsize
\vskip\medskipamount 
\leaders\vrule width \textwidth\vskip0.4pt 
\vskip\medskipamount 
\nointerlineskip

\section{\label{sec:MATLABcode}MATLAB code}

\subsection{\label{sec:boxminus}boxminus.m}

\footnotesize
\begin{verbatim}
function y = boxminus(A1,A2);

y = max(A1-A2,0)+max(A2-A1,0)';
\end{verbatim}
\normalsize
\vskip\medskipamount 
\leaders\vrule width \textwidth\vskip0.4pt 
\vskip\medskipamount 
\nointerlineskip

\subsection{\label{sec:cequiverentropy}cequiverentropy.m}

\footnotesize
\begin{verbatim}
function y = cequiverentropy(A);

% Computes the entropy (in nats) of a componentwise Eulerian quiver with
% adjacency matrix A

%% Basic error checking
if size(A,1)-size(A,2)
    error('A is not square');
end

deg1 = sum(A,1);
deg2 = sum(A,2)';
if any(deg1-deg2)
    error(['sum(A,1) not equal to sum(A,2),'... 
        'so A is not componentwise Eulerian']);
end

if ~any(any(A))
    y = 0; return;
end

%% Shrink A by retaining only nontrivial vertices
ind = find(deg1);   % = find(deg2);
A = A(ind,ind);

%% Find (strongly) connected components of A
scc = strongconcoms(A);
nscc = max(scc);

%% Compute H(A(find(scc==j),find(scc==j)))
H = 0;
for j = 1:nscc
    sccj = find(scc==j);
    Aj = A(sccj,sccj);
    Hj = equiverentropy(Aj);
    H = H+Hj;
end

y = H;
\end{verbatim}
\normalsize
\vskip\medskipamount 
\leaders\vrule width \textwidth\vskip0.4pt 
\vskip\medskipamount 
\nointerlineskip

\subsection{\label{sec:concatenateradixwordquiver}concatenateradixwordquiver.m}

\footnotesize
\begin{verbatim}
function y = concatenateradixwordquiver(u,v,Au,Av,alpha);

% Given two compatible outputs of radixwordquiver, this RAPIDLY produces 
% the quiver corresponding to the concatenation of the underlying words. 
% That is, given words u, v with corresponding radix word quiver matrices
% Au, Av, the RWQ matrix for uv is Auv = Au+Av+B, where B corresponds to 
% the deletion of 2k edges and the insertion of 2k (typically different) 
% edges. Note that this is symmetric because the cyclic word uv equals the 
% cyclic word vu. 
% alpha is the alphabet used to construct Au and Av.
% Example:
% i1 = 56; i2 = 78;
% temp = concatenateradixwordquiver(fp.seq{i1},fp.seq{i2},...
%     rwq{3}{i1},rwq{3}{i2},'ACGTN');
% temp2 = radixwordquiver([fp.seq{i1},fp.seq{i2}],3,'ACGTN');
% temp-temp2 % Should be zero (as when tested at time of writing)

%% Rudimentary error checking
s = size(Au);
n = length(alpha);
k = fix(log(s(1))/log(n));
if any(s-size(Av))
    error('incompatible sizes');
elseif s(1)-s(2)
    error('not square');
elseif abs(k-log(s(1))/log(n)) > 10^-6
    error('k not an integer');
end

%% Other preliminaries
nary = n.^((k-1):-1:0)';
Lu = length(u);
Lv = length(v);
% Map letters to 0:(n-1)
U = zeros(size(u));
V = zeros(size(v));
for j = 1:n
    U(u==alpha(j))=j-1;
    V(v==alpha(j))=j-1;
end
UU = [U,U];
VV = [V,V];
UV = [U,V];
VU = [V,U];
% Initialize sparse matrix
B = sparse(n^k,n^k);

%% Deletions
for j = 1:k
    del01 = UU((Lu-k+j):(Lu-1+j))*nary;
    del11 = UU((Lu-k+j+1):(Lu+j))*nary;
    B(del01+1,del11+1) = B(del01+1,del11+1)-1; 
    del02 = VV((Lv-k+j):(Lv-1+j))*nary;
    del12 = VV((Lv-k+j+1):(Lv+j))*nary;
    B(del02+1,del12+1) = B(del02+1,del12+1)-1; 
end

%% Insertions
for j = 1:k
    ins01 = UV((Lu-k+j):(Lu-1+j))*nary;
    ins11 = UV((Lu-k+j+1):(Lu+j))*nary;
    B(ins01+1,ins11+1) = B(ins01+1,ins11+1)+1; 
    ins02 = VU((Lv-k+j):(Lv-1+j))*nary;
    ins12 = VU((Lv-k+j+1):(Lv+j))*nary;
    B(ins02+1,ins12+1) = B(ins02+1,ins12+1)+1; 
end

%% Output
y = Au+Av+B;
\end{verbatim}
\normalsize
\vskip\medskipamount 
\leaders\vrule width \textwidth\vskip0.4pt 
\vskip\medskipamount 
\nointerlineskip

\subsection{\label{sec:equiverentropy}equiverentropy.m}

\footnotesize
\begin{verbatim}
function y = equiverentropy(A);

% Computes the entropy (in nats) of an Eulerian quiver with adjacency 
% matrix A. 

% Basic care has been exercised to ensure that most of the computation time 
% is spent on the fundamental determinant evaluation instead of, e.g., 
% computing a GCD or the sum of log-factorial terms. There is probably
% still room for improvement though.

%% Force A to be sparse to enable the sparse LU factorization below
A = sparse(A);

%% Basic error checking
if size(A,1)-size(A,2)
    error('A not square');
end
deg1 = sum(A,1);
deg2 = sum(A,2)';
if any(deg1-deg2)
    error('A not Eulerian');
end
if any(any(A<0))
    warning('negative entry'); y = NaN; return;
end
if ~any(any(A))
    y = 0; return;
end

%% Shrink A by retaining only nontrivial vertices
ind = find(deg1);   % = find(deg2);
A = A(ind,ind);

%% Build table for quicksumlogfactorial function below
% The largest degree governs the required table size
logfac = cumsum(log(1:max(deg1)));

%% Find g = gcd(A) non-recursively (for speed) and construct its divisors
% This approach is much faster in MATLAB than a recursive GCD, let alone
% one that doesn't restrict consideration to unique nonzero entries of A.
% In, e.g., C this would not be the correct approach
[J,K] = find(A);
g = 1;
if numel(J)
    % Form vertical array of relevant entries
    AJK = A(sub2ind(size(A),J,K));
    % Only consider the unique values in AJK (note that these are sorted
    % from smallest to largest). NB. This approach appears to be faster
    % than, e.g., omitting [J,K] = find(A), changing "if numel(J)" to 
    % "if any(A)" and setting AJK = setdiff(unique(A),0);
    AJK = unique(AJK);
    % The gcd must be a divisor of AJK(1)--so first form all the divisors
    divAJK1 = divisors(AJK(1));
    % Now just look for any remainders under trial division of the entries 
    % of A by each divisor
    for k = 1:numel(divAJK1)
        if ~mod(AJK,divAJK1(k))
            g = divAJK1(k); 
        end
    end
end
div = divisors(g);
nd = numel(div);

%% Totient of each divisor
phi = ones(1,nd);
for k = 1:nd
    d = div(k);
    fd = factor(d);
    ufd = unique(fd);
    nufd = length(ufd);
    phi(k) = 1;
    for m = 1:nufd
        nm = sum(fd==ufd(m));
        phi(k) = phi(k)*max((ufd(m)^(nm-1))*(ufd(m)-1),1);
    end
end

%% Compute the entropy
H = 0;  % The initialization to 0 vs -Inf is ameliorated at the very end
for k = 1:nd
    d = div(k);
    Ad = A/d;
    degd = sum(Ad,2)';	% = sum(Ad,1)
    L = diag(degd)-Ad;
    L2 = L(2:end,2:end);
    %% Compute log(det(L2)) with a modicum of care
    % A naive yet nontrivial approach would be to compute
    %    logt = real(trace(logm(full(L2))));
    % But this is bad: logm requires a full matrix and it's a
    % computationally nasty way to do things--though not quite as bad as
    %    logt = log(det(L2));
    % We can do better with a sparse LU decomposition
    %    [LL,UL,PL,QL,RL] = lu(L2);
    %    logt = sum(log(abs(diag(UL))))+sum(log(abs(diag(RL))));
    % though on at least one occasion this gave the error message 
    % "Sparse LU failed." This error was not found on the internet and no
    % real attempt was made to understand it--instead we use the
    % perfunctory remedy of a try-catch block to deal with it below using
    % the alternative of the Q-less sparse QR decomposition, e.g.
    %	 RL = qr(L2);
    %	 logt = sum(log(abs(diag(RL))));
    % which folklore suggests is more stable--though testing indicates that
    % this is also much slower.
    %
    % We could go much deeper down the rabbit hole, exploiting the sparsity
    % and integrality of L with great sophistication by, e.g. using 
    % Chinese remaindering/lifting and/or Smith normal forms). A partial
    % bibliography of relevant references follows:
    % \bibitem{PanYuStewart}Pan, V. Y., Yu, Y., and Stewart, C. ``Algebraic
    % and numerical techniques for the computation of matrix
    % determinants.'' \emph{Comp. Math. Appl.} {\bf 34}, 43 (1997).
    % \bibitem{KaltofenVillard}Kaltofen, E. and Villard, G. ``Computing the
    % sign or the value of the determinant of an integer matrix, a
    % complexity survey.'' \emph{J. Comp. Appl. Math.} {\bf 162}, 133
    % (2004).
    % \bibitem{Ogita}Ogita, T. ``Exact determinant of integer matrices.''
    % \emph{Proc. 4th. Int. Workshop on Reliable Engineering Computing.}
    % (2010).
    % \bibitem{ElsheikhEtAl}Elsheikh, M. \emph{et al.} ``Fast computation
    % of Smith forms of sparse matrices over local rings.'' \emph{ISAAC}
    % (2012).
    % \bibitem{EberlyEtAl}Eberly, W. \emph{et al.} ``Faster inversion and
    % other black box matrix computations using efficient block
    % projections.'' \emph{ISAAC} (2007).
    % \bibitem{DumasSaundersVillard}Dumas, J.-G., Saunders, B. D., and
    % Villard, G. ``On efficient sparse integer matrix Smith normal form
    % computations.'' \emph{J. Symb. Comp.} {\bf 32}, 71 (2001).  
    try 
        % NB. It may be advantageous to permute L2, but as yet we don't
        [LL,UL,PL,QL,RL] = lu(L2);
        logt = sum(log(abs(diag(UL))))+sum(log(abs(diag(RL))));
    catch err
        if (strcmp(err.identifier,'MATLAB:sparseLUfactor:internalError'))
            % Try the supposedly more stable but slower Q-less QR
            RL = qr(L2);
            logt = sum(log(abs(diag(RL))));
        else
            rethrow(err);
        end
    end    
    %%
    logc = logt+quicksumlogfactorial(degd-1,logfac);
    logfAd = quicksumlogfactorial(Ad,logfac);
    logterm = log(phi(k))+logc-log(d)-logfAd;
    H = logplus(H,logterm);
end

%% Output
% "H = log(exp(H)-1)" here comes from initialization to 0 vs -Inf (see the
% start of the "Compute the entropy" code cell/section) 
y = H+log(1-exp(-H));

end % END MAIN FUNCTION

%% LOCAL FUNCTIONS BELOW

%%
function y = divisors(x);

% Returns the divisors of x

fx = factor(x);
nf = numel(fx);
mask = zeros(2^nf,nf);
temp = dec2bin(0:((2^nf)-1),nf);
div0 = zeros(2^nf,1);
for j = 1:2^nf
    for k = 1:nf
        mask(j,k) = str2num(temp(j,k));
    end
    div0(j) = prod(max(mask(j,:).*fx,1));
end
y = unique(div0)';

end

%%
function y = logplus(loga,logb);

% Given log(a) = loga and log(b) = logb, returns y = log(a+b). Naively, 
% y = loga+log(1+exp(logb-loga));
% However, we need to take a little care to avoid under/overflow.

X = max(loga,logb);
x = min(loga,logb);
y = X+log(1+exp(x-X));

end

%%
function y = quicksumlogfactorial(M,logfac);

% Given a nonnegative integer array M, compute the sum of the logarithms of
% factorials in M in a careful way using the precomputed table 
% logfac = cumsum(log(1:mM)), where mM = max(...max(M)...)

x = 0;
M = M(find(M>1));	% find(M>1) is (MUCH) better to use than find(M)
for j = 1:numel(M)
    x = x+logfac(M(j));
end
y = x;

end
\end{verbatim}
\normalsize
\vskip\medskipamount 
\leaders\vrule width \textwidth\vskip0.4pt 
\vskip\medskipamount 
\nointerlineskip

\subsection{\label{sec:fastaparse}fastaparse.m}

\footnotesize
\begin{verbatim}
function y = fastaparse(fastafile);

% Parses FASTA format text file. The (canonical) example used here is a
% file containing 616 complete mammalian mitochondrial genomes accessed 
% 20140124 from the URL
% http://www.ncbi.nlm.nih.gov/nuccore/?term=mitochondrion[Definition]+AND+"
% complete+genome"[Definition]+AND+srcdb_refseq[Properties]+AND+Mammalia[Or
% ganism]
% 
% A smaller example is a file containing 109 complete primate genomes
% accessed 20140205 from the URL
% http://www.ncbi.nlm.nih.gov/nuccore/?term=mitochondrion[Definition]+AND+"
% complete+genome"[Definition]+AND+srcdb_refseq[Properties]+AND+Primates[Or
% ganism]

% Example usage for mtDNA FASTA files:
% fp = fastaparse(fastafile);
% for j = 1:length(fp.desc)
%     i1 = max(strfind(fp.desc{j},'|'))+2; 
%     i2 = min(strfind(fp.desc{j},'mitochondrion, complete genome'))-1;
%     desc{j} = fp.desc{j}(i1:i2); 
%     [genus{j},remain{j}] = strtok(desc{j},' '); 
%     [species{j},remain{j}] = strtok(remain{j},' ');
%     seq{j} = fp.seq{j};
% end

fasta = fileread(fastafile);

lb = linebreak(fasta);

L = length(lb);

% Get line numbers for description lines
dline = find(cellfun(@numel,strfind(lb,'>')));
N = length(dline);
% Phantom line for convenience
dline = [dline,L];

% Assemble components into sequences and their descriptions
for j = 1:N
    n = dline(j);
    s.desc{j} = strtrim(lb{n});
    n = n+1;
    s.seq{j} = '';
    while n < dline(j+1)
        s.seq{j} = [s.seq{j},strtrim(lb{n})];
        n = n+1;
    end
    %% FOR CONVENIENCE replace any non-ACGT characters with 'N'
    s.seq{j} = regexprep(s.seq{j},'[^ACGT]','N');
end

y = s;
\end{verbatim}
\normalsize
\vskip\medskipamount 
\leaders\vrule width \textwidth\vskip0.4pt 
\vskip\medskipamount 
\nointerlineskip

\subsection{\label{sec:genbankorganism}genbankorganism.m}

\footnotesize
\begin{verbatim}
function y = genbankorganism(genbankfile);

% Used to extract organism data from a Genbank file

gb = fileread(genbankfile);

lb = linebreak(gb);

L = length(lb);

% Get line numbers for ORGANISM lines
orgline = find(cellfun(@numel,strfind(lb,'  ORGANISM  ')));
% Get line numbers for REFERENCE lines
refline = find(cellfun(@numel,strfind(lb,'REFERENCE   ')));

N = length(orgline);
for j = 1:N
    % Get the first REFERENCE line after each ORGANISM line
    nextrefline = refline(find(refline>orgline(j),1));
    lines = (orgline(j)+1):(nextrefline-1);
    tax{j} = '';
    for k = 1:numel(lines)
        tax{j} = [tax{j},strtrim(lb{lines(k)})];
    end
    tax{j} = strrep(tax{j},' ','');
    tax{j} = strrep(tax{j},'.','');
    remain = tax{j};
    k = 1;
    while remain
        [taxa{j}{k},remain] = strtok(remain,';');
        k = k+1;
    end
end

y = taxa;
\end{verbatim}
\normalsize
\vskip\medskipamount 
\leaders\vrule width \textwidth\vskip0.4pt 
\vskip\medskipamount 
\nointerlineskip

\subsection{\label{sec:linebreak}linebreak.m}

\footnotesize
\begin{verbatim}
function y = linebreak(x);

% Break a string x into nontrivial lines according to return characters. 
% Newlines are replaced with return characters beforehand. 
% Recall that \n = char(10) and \r = char(13)

x = strrep(x,char(10),char(13));

tworets = strfind(x,char([13 13]));
while tworets
    x = strrep(x,char([13 13]),char(13));
    tworets = strfind(x,char([13 13]));
end

rets = find(x==char(13));
lr = length(rets);

if lr == 0
    lb{1} = x;
    lb{2} = '';
else
    lb{1} = x(1:rets(1));
    for j = 2:lr
        lb{j} = x((rets(j-1)+1):rets(j));
    end
    lb{lr+1} = x(rets(end)+1:end);
end

y = lb;

% NB. Reassembly is easy, but included explicitly here in a comment for
% convenience. Below, if lb = linebreak(x), then x2 = x:
% x2 = ''; for j = 1:length(lb), x2 = [x2,lb{j}]; end
\end{verbatim}
\normalsize
\vskip\medskipamount 
\leaders\vrule width \textwidth\vskip0.4pt 
\vskip\medskipamount 
\nointerlineskip

\subsection{\label{sec:phylotree}phylotree.m}

\footnotesize
\begin{verbatim}
function y = phylotree(tree,labs,taxa);

% Automatically produce a rooted phylogenetic tree.
% tree should be obtained from linkage of data with labels labs.
% taxa = genbankorganism(genbankfile);

%% Preliminary plot
figure;
[H,T,outperm] = dendrogram(tree,0,'orientation','right','labels',labs);
set(H,'Color','k');
hold;
% Plot geometry
DAR = get(gca,'DataAspectRatio');
xyR = DAR(1)/DAR(2);
xL = get(gca,'XLim');

%% Get inverse of outperm
N = length(outperm);
temp = sortrows([outperm',(1:N)']);
invperm = temp(:,2)';

%% Rudimentary error checking
if N-numel(labs)
    error('wrong number of labels');
end

%% Get children of jth internal node
% Nodes are 1:N for leaves and (N+1):(2*N-1) for internals, from bottom up. 
% I.e., the shortest branches come from node N+1, etc.
for j = 1:N-1
    children{j} = tree(j,1:2);
    while any(children{j}>N)
        % Find next child with children (an "adult")
        ind = find(children{j}>N,1,'first');
        % Remove the adult and add its children
        adult = children{j}(ind);
        children{j} = union(setdiff(children{j},adult),tree(adult-N,1:2));
    end    
end

%% For each internal node, find two key cell arrays of words
% A: the intersection of taxa words under the node
% B: the union of taxa words NOT under the node
% Along the way, find the union alltaxa of all taxa words for the tree
alltaxa = {};
% This loop builds A and alltaxa
for j = (N-1):-1:1
    t = taxa{children{j}(1)};
    alltaxa = union(alltaxa,t);
    A{j} = t;
    for k = 2:length(children{j})
        t = taxa{children{j}(k)};
        alltaxa = union(alltaxa,t);
        A{j} = intersect(A{j},t);
        %B0{j} = union(B0{j},t);
    end
end
% % This loop builds B
for j = (N-1):-1:1
    otherchildren = setdiff(1:N,children{j});
    B{j} = {};
    for k = 1:length(otherchildren)
        B{j} = union(B{j},taxa{otherchildren(k)});
    end
end

%% For each internal node, its label is the "highest-ranking" member of A\B
for j = (N-1):-1:1
    % The first child is as good as any to use...
    t = taxa{children{j}(1)};
    C = setdiff(A{j},B{j});
    label{j} = '';
    for k = length(t):-1:1
        if any(strcmp(t{k},C))
            label{j} = t{k};
        end
    end
end

%% Add labels
dx1 = .005*(xL(2)-xL(1));
for j = 1:N-1
    if length(label{j})
        x1 = tree(j,3)+2*dx1;
        x2 = mean(invperm(children{j}));
        text(x1,x2,label{j},'FontUnits','normalized','FontSize',1/N);
        x2max = max(invperm(children{j}));
        x2min = min(invperm(children{j}));
        x01 = x1-dx1;
        plot([x01,x01],[x2min,x2max],'r');
    end
end

%% Axes ticks/etc.
set(gca,'FontUnits','normalized','FontSize',1/N);
set(gca,'TickLength',[0 0]);

y = label;
\end{verbatim}
\normalsize
\vskip\medskipamount 
\leaders\vrule width \textwidth\vskip0.4pt 
\vskip\medskipamount 
\nointerlineskip

\subsection{\label{sec:radixwordquiver}radixwordquiver.m}

\footnotesize
\begin{verbatim}
function y = radixwordquiver(w,k,alpha);

% Constructs a de Bruijn quiver of order k for a word w over the alphabet
% alpha, whose order is presumed to induce the lex order, which is used for
% the adjacency matrix A that is actually returned.
% Ex. y = radixwordquiver('GGATTAATGACTAATCAGC',1,'ACGT');
% NB. Instead of 'ACGT' for the last argument, it might be necessary to
% use, e.g. 'ACGTN', cf. fastaparse.

%% 
if numel(setdiff(unique(w),alpha))
    setdiff(unique(w),alpha)
    error('wrong alphabet');
end
% Alphabet radix
n = length(alpha);
if n^k > 2^64
    error('n^k > 2^64; try wordquiver instead');
end

%% Preliminaries
L = length(w);
% Cheap proxy for cyclic word
w = [w,w(1:k)];
% Map letters to 0:(n-1)
W = zeros(size(w));
for j = 1:n
    W(w==alpha(j))=j-1;
end
% Initialize sparse matrix
A = sparse(n^k,n^k);

%% Initialization
% Initial k-gram
J0 = 1;
J1 = k;
kg = W(J0:J1);
% ind1 is the index (starting from zero) for the initial k-gram
nary = n.^((k-1):-1:0)';
ind1 = kg*nary;

%% Main loop
for j = 1:L
    ind0 = ind1;
    % Get upcoming k-gram
    J0 = j+1;
    J1 = j+k;
    kg = W(J0:J1);
    ind1 = kg*nary;
    % Update the adjacency matrix
    A(ind0+1,ind1+1) = A(ind0+1,ind1+1)+1;
end

%% Output
y = A;
\end{verbatim}
\normalsize
\vskip\medskipamount 
\leaders\vrule width \textwidth\vskip0.4pt 
\vskip\medskipamount 
\nointerlineskip

\subsection{\label{sec:strongconcoms}strongconcoms.m}

\footnotesize
\begin{verbatim}
function y = strongconcoms(A);

% Strongly connected components of a quiver with adjacency matrix A using
% the Dulmage-Mendelsohn decomposition

[n,n2] = size(A);

%% Basic error check
if n-n2
    error('A is not square');
end

%% Set zero entries of diagonal to unity so that dmperm works correctly
A = A-spdiags(diag(A),0,n,n)+speye(n);

%% Dulmage-Mendelsohn decomposition
[p,q,r,s,cc,rr] = dmperm(A);
if any(p-q) | any(r-s)
    error('p-q or r-s have a nonzero entry');
end

%% Assign strongly connected components
scc = zeros(1,n);
for j = 1:(length(r)-1)
    scc(p(r(j):(r(j+1)-1))) = j;
end

%% Output
y = scc;
\end{verbatim}
\normalsize
\vskip\medskipamount 
\leaders\vrule width \textwidth\vskip0.4pt 
\vskip\medskipamount 
\nointerlineskip

\subsection{\label{sec:wordquiver}wordquiver.m}

\footnotesize
\begin{verbatim}
function y = wordquiver(w,k);

% Constructs a de Bruijn quiver of order k for a word w. Returns a struct
% with fields for the k-grams and the adjacency matrix with corresponding
% rows/columns.

%% Preliminaries
L = length(w);
% Cheap proxy for cyclic word
w = [w,w(1:k)];
% Initialize cell array for k-grams
kgs = {};
% Initialize sparse matrix
A = sparse(1,1);

%% Initialization
% Initial k-gram
J0 = 1;
J1 = k;
kg1 = w(J0:J1);
% Initialize cell array for k-grams
kgs{1} = kg1;
% ind1 == 1 is the index for the initial k-gram
ind1 = find(strcmp(kg1,kgs));

%% Main loop
for j = 1:L
    kg0 = kg1;
    ind0 = ind1;
    % Get upcoming k-gram
    J0 = j+1;
    J1 = j+k;
    kg1 = w(J0:J1);
    % See if we've encountered the upcoming k-gram before
    ind1 = find(strcmp(kg1,kgs));
    nind1 = numel(ind1);
    if nind1 == 1   % We've seen the upcoming k-gram before...
        % ...so do nothing
    elseif nind1 == 0   % We haven't seen it yet, so include it...
        ind1 = length(kgs)+1;
        kgs{ind1} = kg1;
        % ...and appropriately expand the adjacency matrix
        A(ind1,ind1) = 0;   
    else    % Something is amiss
        error('more than one index');
    end
    % Update the adjacency matrix
    A(ind0,ind1) = A(ind0,ind1)+1;
end

%% Output
y.kg = kgs;
y.A = A;
\end{verbatim}
\normalsize
\vskip\medskipamount 
\leaders\vrule width \textwidth\vskip0.4pt 
\vskip\medskipamount 
\nointerlineskip

\end{document}